\documentclass[a4paper,preprintnumbers,floatfix,twocolumn,aps,prb,unsortedaddress,superscriptaddress]{revtex4-2}

\usepackage[para]{threeparttable} 
\usepackage{amsmath}
\usepackage{graphicx} 
\usepackage{booktabs}
\usepackage{subcaption}
\usepackage{url}
\usepackage{hyperref}
\usepackage{miller}
\usepackage{bm}
\usepackage{enumitem}
\usepackage{color}

\usepackage[paperwidth=210mm,paperheight=297mm,centering,hmargin=1.7cm,vmargin=2.5cm]{geometry}

\usepackage[utf8]{inputenc}

\begin{document}

\title{Multiscale machine-learning interatomic potentials for ferromagnetic and liquid iron}

\author{J. Byggmästar}
\thanks{Corresponding author}
\email{jesper.byggmastar@helsinki.fi}
\affiliation{Department of Physics, P.O. Box 43, FI-00014 University of Helsinki, Finland}
\author{G. Nikoulis}
\affiliation{Department of Physics, P.O. Box 43, FI-00014 University of Helsinki, Finland}
\author{A. Fellman}
\affiliation{Department of Physics, P.O. Box 43, FI-00014 University of Helsinki, Finland}
\author{F. Granberg}
\affiliation{Department of Physics, P.O. Box 43, FI-00014 University of Helsinki, Finland}
\author{F. Djurabekova}
\affiliation{Department of Physics, P.O. Box 43, FI-00014 University of Helsinki, Finland}
\affiliation{Helsinki Institute of Physics, Helsinki, Finland}
\author{K. Nordlund}
\affiliation{Department of Physics, P.O. Box 43, FI-00014 University of Helsinki, Finland}

\date{\today}

\begin{abstract}
{We develop and compare four interatomic potentials for iron: a simple machine-learned embedded atom method (EAM) potential, a potential with machine-learned two- and three-body-dependent terms, a potential with machine-learned EAM and three-body terms, and a Gaussian approximation potential with the SOAP descriptor. All potentials are trained to the same diverse database of body-centered cubic and liquid structures computed with density functional theory. The four presented potentials represent different levels of complexity and span three orders of magnitude in computational cost. The first three potentials are tabulated and evaluated efficiently using cubic spline interpolations, while the fourth one is implemented without additional optimization.  We compare and discuss the advantages of each implementation, transferability and applicability in terms of the balance between required accuracy versus computational cost.
}
\end{abstract}

\maketitle

\section{Introduction}
\label{sec:intro}

As the principal component of all steels and hence arguably the most important metal for industrial and structural applications, iron is one of the most intensely modelled materials. Its magnetic nature is the source of many interesting properties that separate iron from other body-centered cubic metals, such as its high-temperature phase transitions~\cite{hasegawa_microscopic_1983} and the exotic landscape of radiation-induced defects~\cite{marinica_irradiation-induced_2012,dudarev_effect_2008,terentyev_self-trapped_2008}. This makes developing accurate interatomic potentials for large-scale atomistic modeling of iron challenging. Consequently, a large number of interatomic potentials have been developed in the last decades, targeting different key properties. Most existing potentials are traditional parametric analytical potentials, like embedded atom method (EAM) potentials~\cite{ackland_computer_1997,mendelev_development_2003,ackland_development_2004,malerba_comparison_2010,dudarev_magnetic_2005,zhou_misfit-energy-increasing_2004,olsson_semi-empirical_2009,chiesa_optimization_2011,alexander_interatomic_2020}, angular-dependent modified EAM potentials~\cite{lee_second_2001,asadi_quantitative_2015,etesami_molecular_2018,starikov_angular-dependent_2021}, and Tersoff-like or magnetic analytical bond-order potentials (ABOP)~\cite{muller_analytic_2007,byggmastar_dynamical_2020,mrovec_magnetic_2011,lin_bond-order_2016-1}. Even though these potentials have been very successful in describing most properties of iron, recent machine-learning potentials have provided a new level of accuracy for e.g. thermal, defect, and screw dislocation properties~\cite{dragoni_achieving_2018,mori_neural_2020,goryaeva_efficient_2021}. Very recently, there has also been progress in explicitly including spins in machine-learning potentials~\cite{novikov_machine-learning_2020,eckhoff_high-dimensional_2021} or coupling a machine-learning potential to a spin model~\cite{nikolov_data-driven_2021} to quantitatively reproduce magnetism in iron and other materials.

Exploiting machine learning (ML) is now rapidly becoming routine when constructing and fitting interatomic potentials. A growing number of different ML frameworks and descriptors have been developed in what is now an extremely active research field~\cite{behler_perspective_2016,mueller_machine_2020}. Potentials using different underlying ML methods (artificial neural networks~\cite{behler_generalized_2007}, kernel regression~\cite{bartok_gaussian_2010}, linear regression~\cite{shapeev_moment_2016,thompson_spectral_2015}, and deep learning~\cite{zhang_deep_2018}) have all demonstrated near-quantum accuracy for all classes of materials~\cite{zuo_performance_2020}.

Despite their success and excellent accuracy, machine-learning potentials have not and will not completely replace traditional parametric interatomic potentials. This is partly because traditional fixed-function potentials offer a transferability that is difficult to achieve with ML potentials, as ML models are inherently poor at extrapolation. Secondly, most ML potentials are computationally much more costly than simple traditional potentials like EAM or Tersoff. The choice of potential type one develops or applies in a simulation should be based on the balance between desired accuracy and the acceptable computational cost. For many molecular dynamics (MD) applications, such as simulating large-scale or long-term irradiation damage~\cite{granberg_molecular_2021}, the computational price of highly accurate machine-learning potentials is simply too high. With this in mind, the aim of this work is to develop, using machine-learning methods, a set of increasingly complex interatomic potentials for iron that provide different levels of accuracy and computational efficiency. In particular, we further develop the methodology of tabulated low-dimensional machine-learning potentials (tabGAP~\cite{byggmastar_modeling_2021}) and show that they can provide an excellent balance between speed, accuracy, and transferability.

\section{Methods}
\label{sec:methods}

\subsection{Gaussian approximation potentials}

All potentials developed here are trained as Gaussian approximation potentials (GAP)~\cite{bartok_gaussian_2010} using different combinations of increasingly complex descriptors. All potentials include a fixed short-range repulsive pair potential ($E_\mathrm{rep}$) appropriate to handle high-energy collision correctly \cite{ziegler_stopping_1985,Nor94b,nordlund_repulsive_1997}, so that the total energy of a system of $N$ atoms is given by
\begin{equation}
    E_\mathrm{tot} = E_\mathrm{rep} + E_\mathrm{ML}.
\end{equation}
The energy (and corresponding forces and stresses) to be machine-learned is hence $E_\mathrm{ML} = E_\mathrm{tot} - E_\mathrm{rep}$, where $E_\mathrm{tot}$ is the total energy of a given structure in the training database computed with density functional theory (DFT). The repulsive pair potential is a screened Coulomb potential fitted to Fe--Fe repulsion and forced to zero by a smooth cutoff function $f_\mathrm{cut} (r_{ij}$) as~\cite{byggmastar_machine-learning_2019}
\begin{equation}
    E_\mathrm{rep} = \sum_{i<j}^N \frac{1}{4\pi \varepsilon_0} \frac{Z_i Z_j e^2}{r_{ij}} \phi_\mathrm{Fe-Fe} (r_{ij}/a) f_\mathrm{cut} (r_{ij}),
\end{equation}
where
\begin{equation}
    a = \frac{0.46848}{Z_i^{0.23} + Z_j^{0.23}},
\end{equation}
as in the universal ZBL potential~\cite{ziegler_stopping_1985}. The cutoff function forces the potential smoothly to zero in the range 1.1--2.2 Å. This is well below the nearest-neighbour distance in bcc (2.45 Å) and hence leaves all near-equilibrium interactions to be machine-learned. The screening function $\phi$ is fitted to reproduce all-electron DFT data for the Fe--Fe dimer repulsion~\cite{nordlund_repulsive_1997} and is given by
\begin{equation}
\begin{aligned}
    \phi_\mathrm{Fe-Fe}(x) & = 0.375708 \exp(-17.2128x) \\
    & + 0.0020925 \exp(-1.297x) \\
    & + 0.622672 \exp(-4.73614x).
\end{aligned}
\end{equation}

The simplest and least accurate potential version, a machine-learned EAM potential, contains two machine-learning terms with pairwise ($E_\mathrm{2b}$) and embedding energy ($E_\mathrm{emb}$) contributions:
\begin{equation}
    E_\mathrm{GAP\text{-}EAM} = E_\mathrm{rep} + E_\mathrm{2b} + E_\mathrm{emb}.
\end{equation}
All machine-learning terms are evaluated using Gaussian process regression as implemented in \textsc{quip}~\cite{QUIP} and part of the GAP framework. Including the EAM-like embedding term has not been done previously in GAP and is explained in detail below. The two-body term can be written
\begin{equation}
    E_\mathrm{2b} = \sum_{i<j}^N \delta^2_\mathrm{2b} \sum_s^{M_\mathrm{2b}} \alpha_{s} K_\mathrm{se} (r_{ij}, r_s),
\end{equation}
where $\delta^2$ is a prefactor, $\alpha_s$ are the regression coefficients, and $K_\mathrm{se}$ is the squared-exponential kernel. The sum runs over a selected (sparsified) subset of known descriptor environments from the training structures (here just the $M_\mathrm{2b}$ interatomic distances $r_s$)~\cite{bartok_gaussian_2015}. The embedding energy is similarly given by
\begin{equation}
    E_\mathrm{emb} = \sum_{i}^N \delta^2_\mathrm{eam} \sum_s^{M_\mathrm{eam}} \alpha_{s} K_\mathrm{se} (\rho_i, \rho_s).
\end{equation}
Here, the descriptor input to the kernel function is the total density contributed by all atoms $j$ in the local atomic environment of $i$, as in a normal EAM potential:
\begin{equation}
    \rho_i = \sum_j^N \varphi_{ij} (r_{ij}).
\end{equation}
The use of an EAM-like density as a simple many-body descriptor for ML potentials was first demonstrated in Ref.~\cite{zeni_gaussian_2020}, although with different expressions for the pairwise and total density. We have implemented several functions for the pairwise density contributions $\varphi_{ij}$. Here, we use the polynomial function
\begin{equation}
    \varphi (r_{ij}) = 
    \begin{cases}
        (-1)^n (r_{ij} - r_\mathrm{cut})^n / r_\mathrm{cut}^n, & r_{ij} \leq r_\mathrm{cut} \\
        0, & r_{ij} > r_\mathrm{cut}, \\
    \end{cases}
\end{equation}
with $n=3$, making the cutoff continuous up to the second derivative. $n=2$ would be the Finnis-Sinclair density function (normalised so that $\varphi (0) = 1$)~\cite{finnis_simple_1984}. $r_\mathrm{cut}$ is the cutoff radius. Since the descriptor is the total pairwise-contributed density, training the GAP-EAM potential effectively means machine-learning the embedding function of an EAM potential together with the pair potential. Note that in normal EAM potentials, the pair potential and the pair density function are often fitted freely using cubic spline functions. The GAP-EAM potential is hence actually less flexible because the pair density function is fixed as a part of the descriptor during the fitting process (although it could in principle be pre-fitted and used as a descriptor). The main practical advantage of the machine-learned embedding term is when combining it with an angular-dependent descriptor as discussed below. The simple GAP-EAM potential is here mainly included for the purpose of comparison with increasingly more flexible machine-learning potentials. It would also be possible to include several embedding terms with different pair density functions, which could be seen as a machine-learning multi-band generalisation of the 2-band EAM potential~\cite{ackland_two-band_2003}. Here, we only use one embedding term and leave investigation of machine-learned multi-band EAM potentials for future work.

We also train a potential with only two- and three-body terms as
\begin{equation}
    E_\mathrm{GAP\text{-}3b} = E_\mathrm{rep} + E_\mathrm{2b} + E_\mathrm{3b}.
\end{equation}
The three-body machine-learning term is
\begin{equation}
    E_\mathrm{3b} = \sum_{i, j < k}^N \delta^2_\mathrm{3b} \sum_s^{M_\mathrm{3b}} \alpha_{s} K_\mathrm{se} (\bm{q}_{ijk}, \bm{q}_s),
\end{equation}
where the descriptor is the three-valued permutation-invariant vector~\cite{bartok_gaussian_2015}
\begin{equation}
    \bm{q}_{ijk} =
    \begin{pmatrix}
    r_{ij} + r_{ik} \\
    (r_{ij} - r_{ik})^2 \\
    r_{jk} \\
    \end{pmatrix}
    f_\mathrm{cut}(r_{ij}) f_\mathrm{cut}(r_{ik}).
\end{equation}

The GAP-EAM potential represents the simplest possible many-body potential and is computationally efficient. However, it contains no angular dependence and can only be expected to work reasonably well for simple metals. In contrast, the GAP-3b potential captures angular information, but the pure three-body dependence is not enough for liquids or amorphous structures, where many-body (higher than three) and proper coordination dependencies are needed to reach good accuracy (as we demonstrate in Sec.~\ref{sec:speed_accuracy}). For a more flexible and generally applicable potential, therefore, it is obvious that both the three-body and the embedding terms should be used as
\begin{equation}
    E_\mathrm{GAP\text{-}3b+EAM} = E_\mathrm{rep} + E_\mathrm{2b} + E_\mathrm{emb} + E_\mathrm{3b}.
\end{equation}
The GAP-3b+EAM potential can be considered a machine-learning alternative to the angular-dependent modified EAM potentials~\cite{baskes_modified_1992,lee_second_2000}.

The final and most complex potential is a typical GAP where the main ingredient is the well-established SOAP descriptor~\cite{bartok_representing_2013}, used here together with the repulsive and machine-learned pair potentials as
\begin{equation}
    E_\mathrm{GAP\text{-}SOAP} = E_\mathrm{rep} + E_\mathrm{2b} + E_\mathrm{SOAP}.
\end{equation}
We refer to this potential as GAP-SOAP. The many-body SOAP term is given by
\begin{equation}
    E_\mathrm{SOAP} = \sum_{i}^N \delta^2_\mathrm{SOAP} \sum_s^{M_\mathrm{SOAP}} \alpha_{s} K_\mathrm{SOAP} (\bm{q}_i, \bm{q}_s),
\end{equation}
where $K_\mathrm{SOAP}$ is the SOAP kernel and $\bm{q}_i$ is the SOAP descriptor vector of the local environment of atom $i$~\cite{bartok_representing_2013}.

\subsection{tabGAP: tabulated Gaussian approximation potentials}

The GAP-EAM, GAP-3b, and GAP-3b+EAM potentials depend only on simple low-dimensional descriptors. Hence, after training they can all be tabulated by mapping the machine-learning energy predictions onto suitable grids~\cite{byggmastar_modeling_2021}. This bypasses the Gaussian process regression sum over the training environments, and yields a significant computational speed-up. The pairwise energies can be trivially tabulated as a function of the interatomic distance $r_{ij}$ and evaluated using a smooth and differentiable one-dimensional cubic spline interpolation. Similarly, the machine-learning embedding term can be tabulated as a function of the total density $\rho$, which in turn is tabulated as a function of $r_{ij}$. The three-body term must be mapped onto a three-dimensional grid and evaluated by a 3D spline interpolation. For this, we choose a grid of $(r_{ij}, r_{ik}, \cos \theta_{ijk})$ points and a 3D cubic spline implementation. $\theta_{ijk}$ is the angle between the $ij$ and $ik$ bonds. With sufficiently dense grids, the interpolation errors are negligible compared to the accuracy of the potential, as demonstrated in Appendix B. Similar tabulation schemes have been developed before for other types of ML potentials~\cite{glielmo_efficient_2018,vandermause_--fly_2020}, although some details differ from our approach.

We refer to the tabulated versions of the low-dimensional GAPs as tabGAPs~\cite{byggmastar_modeling_2021}. With $S$ representing cubic splines, the tabGAP-EAM can be written
\begin{equation}
\begin{aligned}
    & E_\mathrm{GAP\text{-}EAM} \overset{\mathrm{tab.}}{\simeq} E_\mathrm{tabGAP\text{-}EAM} =\\
    & \sum_{i < j}^N S_{\mathrm{rep + 2b}}^\mathrm{1D} (r_{ij}) + S_\mathrm{emb}^\mathrm{1D} \left(\sum_j^N S_\varphi^\mathrm{1D} (r_{ij}) \right),
\end{aligned}
\end{equation}
where the repulsive and ML pair potentials are combined into one spline interpolation. In practice, this represents a normal tabulated EAM potential file and tabGAP-EAM can thus be evaluated normally using any EAM implementation. The tabulated version of GAP-3b becomes
\begin{equation}
\begin{aligned}
    & E_\mathrm{GAP\text{-}3b} \overset{\mathrm{tab.}}{\simeq} E_\mathrm{tabGAP\text{-}3b} =\\
    & \sum_{i < j}^N S_{\mathrm{rep + 2b}}^\mathrm{1D} (r_{ij}) + \sum_{i, j<k}^N S_{ijk}^\mathrm{3D} (r_{ij}, r_{ik}, \cos \theta_{ijk}).
\end{aligned}
\end{equation}
We have implemented this 1D+3D cubic spline interpolation as the \texttt{pair\_style tabgap} for \textsc{lammps}, available from Ref.~\cite{tabgap} along with code for making tabGAP potential files from GAP potential files.

The GAP-3b+EAM becomes the tabulated version
\begin{equation}
\begin{aligned}
    & E_\mathrm{GAP\text{-}3b+EAM} \overset{\mathrm{tab.}}{\simeq} E_\mathrm{tabGAP} =\\
    & \sum_{i < j}^N S_{\mathrm{rep + 2b}}^\mathrm{1D} (r_{ij}) + \sum_{i, j<k}^N S_{ijk}^\mathrm{3D} (r_{ij}, r_{ik}, \cos \theta_{ijk}) \\
    & + S_\mathrm{emb.}^\mathrm{1D} \left(\sum_j^N S_\varphi^\mathrm{1D} (r_{ij}) \right).
\end{aligned}
\end{equation}
For simplicity and because this version is the most accurate and practically useful tabulated potential, we refer to it hereafter simply as the tabGAP. The tabGAP is in practice used with the \texttt{hybrid/overlay} functionality in \textsc{lammps}, combining the \texttt{eam/fs} and \texttt{tabgap} \texttt{pair\_style}s. Note that our original tabGAP for refractory alloys in Ref.~\cite{byggmastar_modeling_2021} used only 2b and 3b terms as in tabGAP-3b.

\subsection{Hyperparameters}

\begin{table}
    \centering
        \caption{Hyperparameters used for the different descriptors: cutoff radius $r_\mathrm{cut}$, width of the cutoff region $r_{\Delta \mathrm{cut}}$, energy prefactor $\delta$, and the number of sparse descriptor environments from the training structures $M$.}
    \begin{tabular}{lllll}
         \toprule
         Descriptor & $r_\mathrm{cut}$ (Å) & $r_{\Delta \mathrm{cut}}$ (Å) & $\delta$ & $M$ \\
         \midrule
         2b & 4.5 & 1.0 & 10 & 20 \\
         EAM & 4.5 & 1.0 & 1.0 & 20 \\
         3b & 3.7 & 0.6 & 1.0 & 500 \\
         SOAP & 4.5 & 1.0 & 2.0 & 3000 \\
         \bottomrule
    \end{tabular}
    \label{tab:hyper}
\end{table}

Table~\ref{tab:hyper} lists the key hyperparameters used  when training the GAPs. The interaction range for all descriptors except three-body includes the third-nearest neighbour atoms in bcc iron. For the three-body descriptor, we found that using a shorter cutoff that only includes second-nearest neighbours provides the best compromise between speed and accuracy. The number of sparse points $M$ were converged to sufficient values by looking at the test errors as functions of $M$. For the energy, force, and virial regularization parameters $\sigma$ used in GAP training~\cite{bartok_machine_2018}, the default values were set to 1 meV/atom, 0.04 eV/Å, and 0.1 eV. For surface structures we used stronger regularization with twice the default values ($2\sigma$) and for liquids $10 \sigma$.

\subsection{Training and testing data}

The training data consists of total energies, forces, and (for some structures) virial stresses computed by density functional theory calculations for 1078 structures containing 1--259 atoms (in total 38613 atoms). The training and testing structures are available from Ref.~\cite{gap-data}. The following types of structures are included in the training data:

\begin{itemize}
    \item Elastically and randomly distorted bcc unit cells.
    \item Single-crystal bcc cells at finite temperatures and a few different volumes.
    \item Single vacancies and clusters up to three vacancies, including various migration path saddle points.
    \item Single self-interstitial atom (SIA) configurations and clusters. The clusters are small (size 2--4) C15 Laves clusters~\cite{marinica_irradiation-induced_2012} and parallel and nonparallel dumbbells. We also included small 1/2\hkl<111> and \hkl<100> dislocation loops.
    \item \hkl(100), \hkl(110), \hkl(111), and \hkl(211) surfaces.
    \item \hkl(100), \hkl(110), and \hkl(211) $\gamma$-surfaces.
    \item 1/2\hkl<111> screw dislocations and 1/2\hkl<111> and \hkl<100> edge dislocations.
    \item Short-range structures for interatomic repulsion, where one interstitial atom is placed randomly in the bcc crystal without relaxation (so that it is relatively close, but not too close, to its neighbour atoms).
    \item Liquids at various densities. To get a reasonable spread from low to high densities, we sampled liquids according to a $\chi^2$-distribution around the density of liquid iron at the melting point and normal pressure from experiments~\cite{assael_reference_2006}.
\end{itemize}

In all structures except the distorted unit cells, the atoms are slightly displaced from the perfect lattice positions to induce non-zero forces and to create unique local atomic environments. This is either done by introducing small random displacements or by picking frames from finite-temperature MD simulations. For many structure types (mainly the liquids and the defect clusters), new structures were created by relaxing or running MD with an early version of the GAP or tabGAP.

During training and when testing and converging hyperparameters, the accuracy was monitored with a test set of crystalline and liquid structures. The test set crystals include bcc lattices with random atom displacements and five 250-atom lattices containing 3--5 randomly inserted Frenkel pairs to test defect properties. The test set also includes five 128-atom liquids.

\subsection{Density functional theory calculations}

All density functional theory calculations are performed with the \textsc{vasp} code~\cite{kresse_ab_1993,kresse_ab_1994,kresse_efficiency_1996,kresse_efficient_1996}. We used the PBE GGA exchange-correlation functional~\cite{perdew_generalized_1996} and the \texttt{Fe\_sv} projector-augmented wave potential~\cite{blochl_projector_1994,kresse_ultrasoft_1999} with 16 valence electrons. The energy cutoff for the plane-wave expansion was 500 eV. The spacing of $k$-points for the Brillouin zone integration was set to a maximum of 0.15 Å$^{-1}$ on $\Gamma$-centered Monkhorst-Pack grids~\cite{monkhorst_special_1976}. A 0.1 eV first-order Methfessel-Paxton smearing~\cite{methfessel_high-precision_1989} was applied. All calculations were done with spin-polarization and collinear magnetic configurations (correspondonding to ferromagnetic Fe in the bcc crystalline structures).

\subsection{Molecular dynamics and statics simulations}
\label{sec:md}

All molecular dynamics and statics simulations for benchmarking the potentials are done using \textsc{lammps}~\cite{plimpton_fast_1995}. Migration barriers are computed with the climbing-image nudged elastic band (NEB) method~\cite{henkelman_climbing_2000} as implemented in \textsc{lammps}.

For most of the test calculations and simulations, to minimise the effort for the slow GAP-SOAP potential, we used the fast tabGAP-EAM to find converged box sizes and simulation times. The thermal expansion for both the bcc and the liquid phase was simulated using 1024 atoms in 20 ps simulations in the $NPT$ ensemble at zero pressure~\cite{nose_molecular_1984,hoover_canonical_1985} and averaging the volume over the last 16 ps. The structure and properties of the liquid were further examined by equilibrating a molten 2000-atom cell at the melting temperature for 100 ps at zero pressure in MD. The final 85 ps were used to get the average potential energy, volume, and radial distribution function. We determined the melting temperature using the solid-liquid interface method in a box of 1372 atoms, i.e. by finding the temperature at which the solid and the liquid phase are in equilibrium~\cite{Mor94}.

All defects were relaxed by minimising the energy and pressure of the system. The single vacancies and divacancies were relaxed in 250-atom bcc lattices, including the NEB calculations. For the single SIA relaxations and NEB calculations we used 1024-atom bcc lattices. The small size 2--4 parallel and nonparallel SIA clusters were inserted and relaxed in 2000-atom bcc lattices. The bigger SIA clusters (up to 100 SIAs) were inserted and relaxed in boxes of 16000 atoms. The dislocation loops were (close-to) circular loops with Burgers vectors \hkl<100> and 1/2\hkl<111>. For sizes below 25 SIAs, we relaxed 1/2\hkl<111> loops with both \hkl{111} and \hkl{110} habit planes and used the lowest-energy configuration for the final formation energy. Overall, the difference in energy was small, so for larger than 25 SIAs we only considered the \hkl{111} plane.

To find low-energy C15 clusters, we carried out growth-annealing simulations with the tabGAP similar to what is described in detail in Ref.~\cite{byggmastar_dynamical_2020}. In short, we started from a stable C15 cluster and inserted random interstitials close to the cluster one by one followed by annealing and final energy minimisation. During annealing, the C15 cluster captures the added interstitial and grows. This is repeated until a desired size is reached. In this way, we grew C15 clusters between sizes 4--40 SIAs, starting from stable size-4, size-11, size-17, and size-30 C15 clusters. We simulated 40 different growth runs for every size range, and extracted the lowest-energy C15 clusters for comparison with the formation energies of dislocation loops in all potentials.

\section{Results and discussion}

\subsection{Accuracy versus speed}
\label{sec:speed_accuracy}

\begin{figure*}
    \centering
    \includegraphics[width=0.8\linewidth]{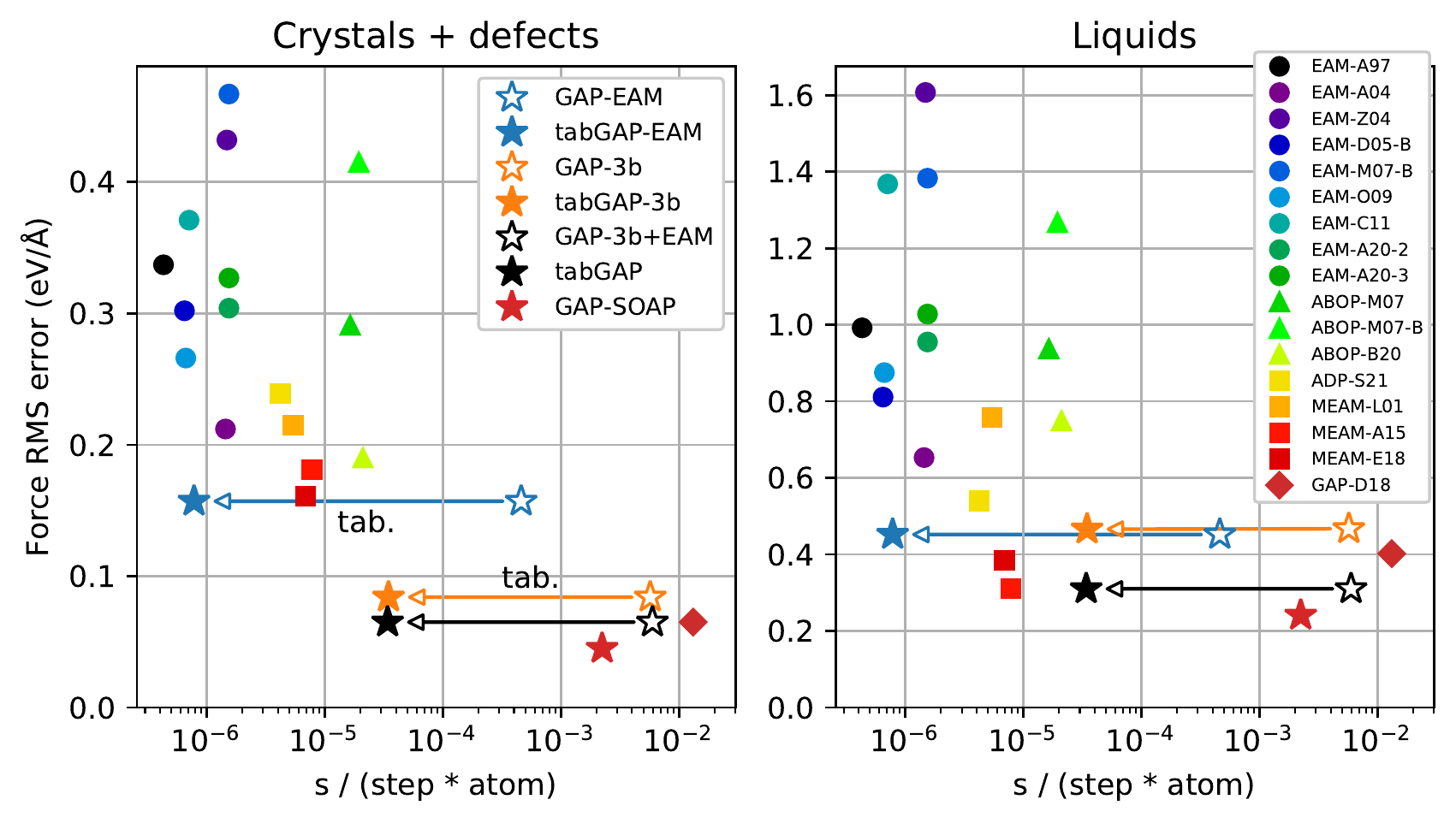}
    \caption{Accuracy versus computational cost of Fe interatomic potentials, shown as root-mean-square deviations from the force components of DFT-computed test structures. The markers distinguish between different types of potentials, with stars indicating the (tab)GAPs developed here. The arrows indicate the speedup of the GAP $\rightarrow$ tabGAP tabulation. The computational cost is tested using a standard \textsc{gcc} v7.5-compiled version of \textsc{lammps} (version 18 Sep 2020) on a single Intel(R) Core(TM) i5-7500 3.40GHz CPU core. The potentials from the literature are named according to type of potential followed by first-author initial and year of publication. In order of appearance, the EAM potentials are from Refs.~\cite{ackland_computer_1997}\cite{ackland_development_2004}\cite{zhou_misfit-energy-increasing_2004}\cite{dudarev_magnetic_2005}\cite{malerba_comparison_2010}\cite{olsson_semi-empirical_2009}\cite{chiesa_optimization_2011}\cite{alexander_interatomic_2020} (EAM-D05-B is the Ref.~\cite{bjorkas_comparative_2007} version of the Ref.~\cite{dudarev_magnetic_2005} EAM, and EAM-M07-B is the Ref.~\cite{byggmastar_effects_2018} version of~\cite{malerba_comparison_2010}), the ABOPs are from Refs.~\cite{muller_analytic_2007}\cite{bjorkas_comparative_2007}\cite{byggmastar_dynamical_2020}, ADP-S21 is from Ref.~\cite{starikov_angular-dependent_2021}, the MEAMs are from Refs.~\cite{lee_second_2001}\cite{asadi_quantitative_2015}\cite{etesami_molecular_2018}, and GAP-D18 from Ref.~\cite{dragoni_achieving_2018}.}
    \label{fig:speed}
\end{figure*}

Fig.~\ref{fig:speed} illustrates the balance between achievable accuracy and computational cost of various types of interatomic potentials for iron. It shows the root-mean-square errors with respect to the DFT force components in the test structures plotted as functions of the computational cost of the potential. EAM potentials are by far the fastest many-body potential, but can only reach a limited level of accuracy. Angular-dependent potentials, like MEAM and ABOP, can be slightly more accurate at the expense of some speed. Fig.~\ref{fig:speed} shows that the machine-learned potentials developed in this work fall into favourable spots in the balance between speed and accuracy compared to existing potentials. However, it should be emphasised that most of the existing potentials have not been force-matched to DFT data, but instead fitted to a mix of experimental and DFT-computed material properties. This makes the comparison with our DFT-computed forces somewhat unfair, but still provides an approximate measure for the performance of different types of potentials. The few notable exceptions that were fitted to DFT forces and liquid properties; EAM-A04, ADP-S21, MEAM-A15, and MEAM-E18, stand out with the lowest errors among the existing potentials in Fig.~\ref{fig:speed}.

Fig.~\ref{fig:speed} also shows the speedup gained by the tabulation of the various low-dimensional GAPs into the corresponding tabGAP versions. The reduction in computational cost is more than two orders of magnitude. For the most accurate and relevant version, the tabGAP, the GAP-3b+EAM $\rightarrow$ tabGAP tabulation provides a 175-times faster potential with no loss in accuracy.

GAP-SOAP is the most accurate of the new potentials, but also by far the slowest (about 65 times slower than the tabGAP). Fig.~\ref{fig:speed} also shows the previous GAP-SOAP (GAP-D18~\cite{dragoni_achieving_2018}), which due to differences in hyperparameters is slower than our GAP-SOAP, but also very accurate. Note that liquids were not included in their training data, which explains the higher test errors for the liquid test set.

\begin{figure}
    \centering
    \includegraphics[width=\linewidth]{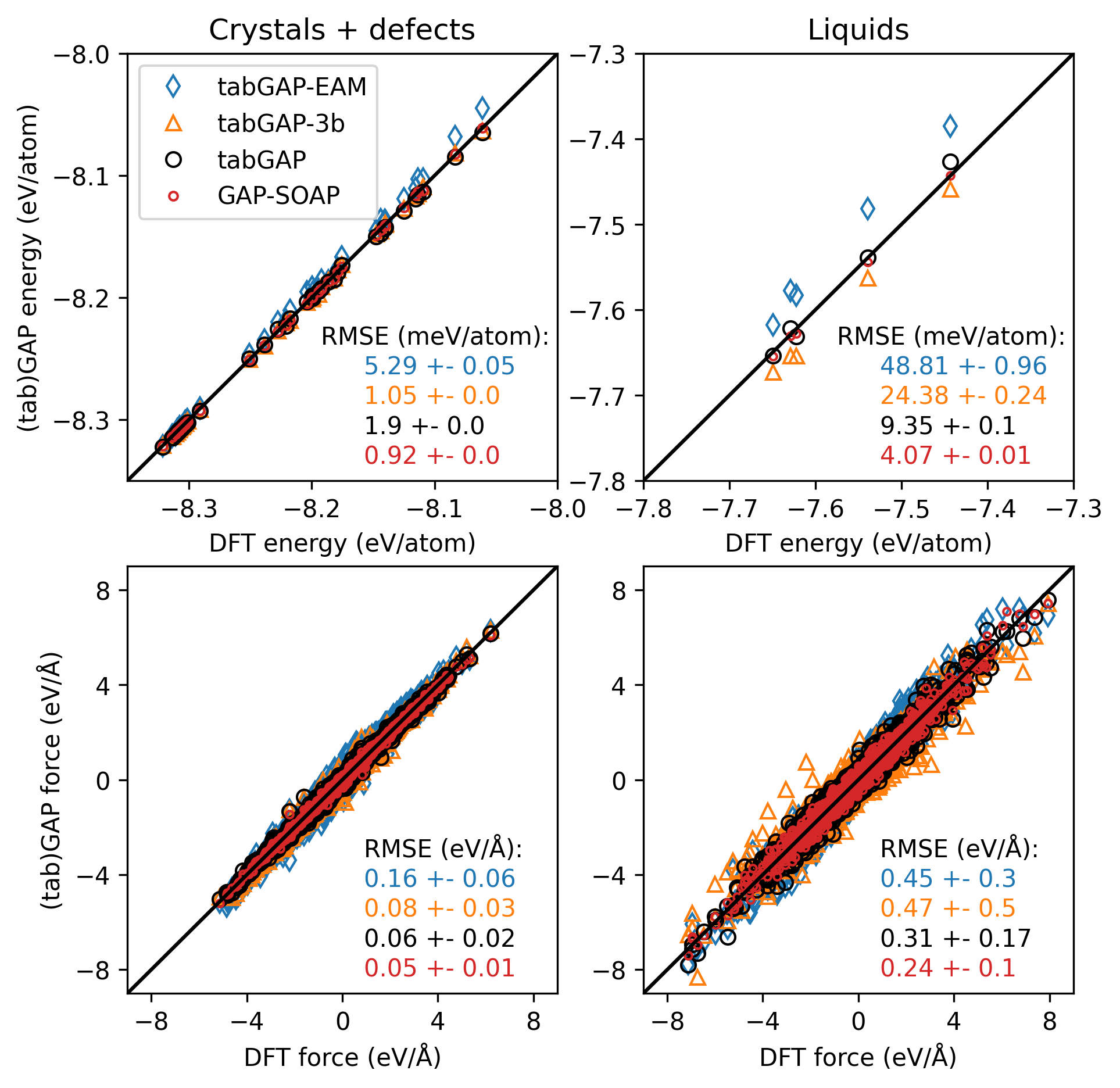}
    \caption{Energy and force components for the test sets, compared between the different potentials and DFT. The root-mean-square errors (RMSE) are listed with the standard deviation of the squared errors as the uncertainty.}
    \label{fig:test}
\end{figure}

\begin{figure}
    \centering
    \includegraphics[width=\linewidth]{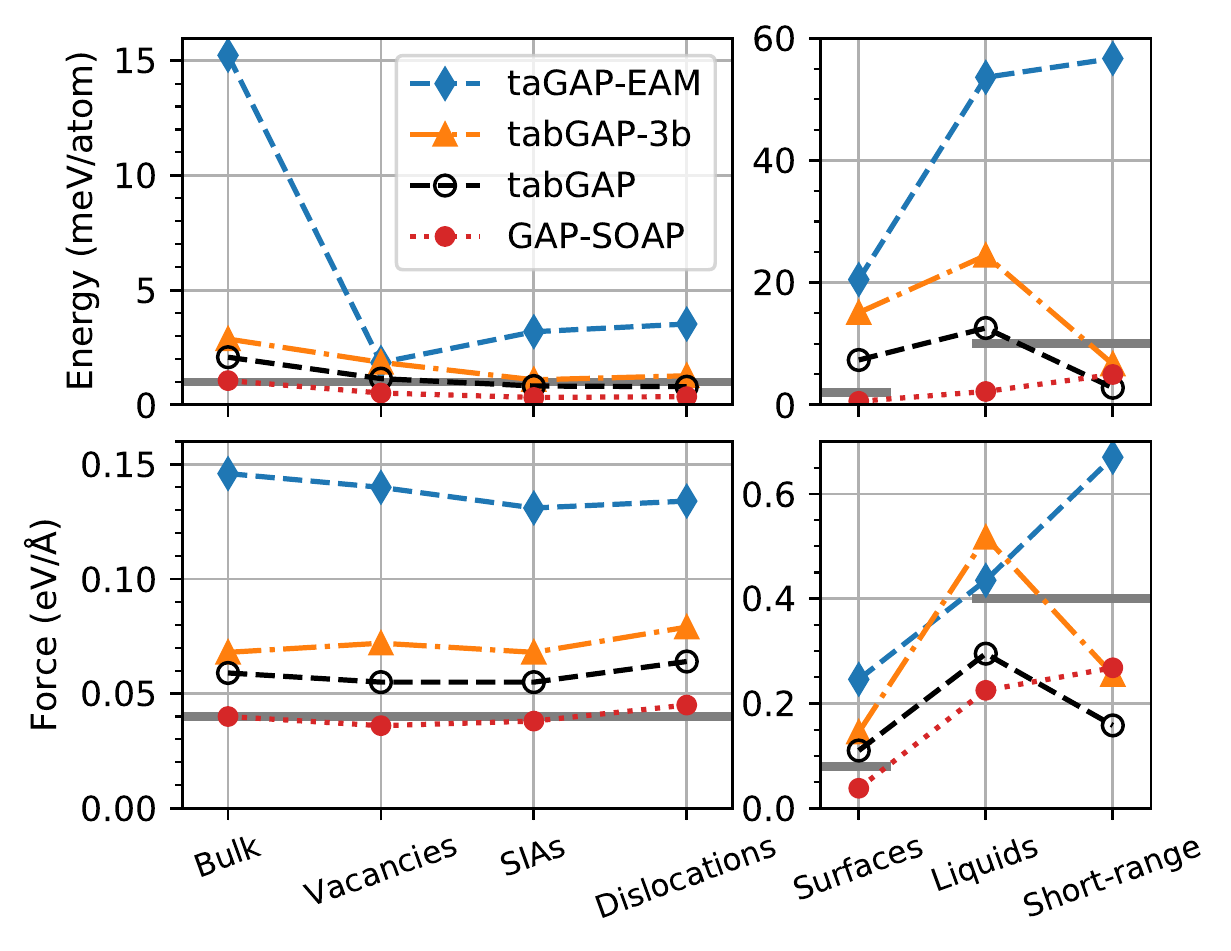}
    \caption{Training errors of the different classes of structures included in the training database, showing the improvement in accuracy with increasingly complex descriptors in the potentials. The grey lines indicate the desired accuracy compared to the DFT data, i.e. the regularisation errors used during the GAP training.}
    \label{fig:train}
\end{figure}

To further examine and compare the training and testing accuracy of the new tabGAPs and GAP-SOAP, Figs.~\ref{fig:test} and \ref{fig:train} show energy and force errors for the test and training data sets. Fig.~\ref{fig:test} is the same test data as in Fig.~\ref{fig:speed}. Comparing the potentials in Figs.~\ref{fig:test} and \ref{fig:train} reveals several noteworthy points. First, the limited flexibility of tabGAP-EAM and tabGAP-3b makes it impossible to reproduce certain structure types with good accuracy. For example, tabGAP-EAM is somewhat overfitted to defects and gives much larger energy errors for simple finite-temperature bulk bcc iron (Fig.~\ref{fig:train}). The accuracy of tabGAP-EAM is overall much worse than the other potentials, which is expected, and can to some degree be accepted given its low computational cost. Second, tabGAP-3b provides very good accuracy for all crystalline structures, but the pure 3-body dependence is clearly insufficient to accurately describe the liquid phase, as seen in both Fig.~\ref{fig:test} and \ref{fig:train}. Using both the 3-body and the EAM descriptor in the tabGAP provides enough flexibility to overcome the above-mentioned issues. Figs.~\ref{fig:test} and \ref{fig:train} show that the accuracy of tabGAP for crystalline structures is still excellent, often very close to GAP-SOAP, and the liquid errors are greatly reduced compared to tabGAP-EAM and tabGAP-3b. The RMS errors for crystalline structures are at most a few meV/atom and around 0.06 eV/Å. For liquids they are reduced to only around 10 meV/atom and 0.3 eV/Å, compared to 20--50 meV/atom and 0.4--0.5 eV/Å for tabGAP-EAM and tabGAP-3b. GAP-SOAP outperforms the tabGAP for all structures slightly, although at a significantly higher computational cost as discussed above.

From here on, we will not include tabGAP-3b in the discussion as it is overall much less accurate than tabGAP but at the same computational cost (the additional cost of the EAM term in tabGAP is negligible compared to the 3-body term).

\subsection{Bulk and surface properties}

\begin{table*}
 \centering
 \caption{Basic bulk, surface, defect, and thermal properties of iron compared between experiments, DFT, and the three potentials. $a$: bcc lattice constant, $B$: bulk modulus and $C_{ij}$: elastic constants. $E_\mathrm{surf}$: surface energy, $E_\mathrm{f}$: formation energies of a single vacancy, dumbbell, octahedral, and tetrahedral interstitials, $E_\mathrm{mig.}$: migration energy, $E_\mathrm{b}$: binding energy, $\alpha_L$: linear thermal expansion coefficient at room temperature, $ T_\mathrm{melt}$: melting point, $\Delta H_\mathrm{melt}$: latent heat, and $\rho_\mathrm{liq.}$: density of liquid iron at the melting point. The experimental structural and elastic properties data from Ref.~\cite{rumble_crc_2019} are measured at room temperature.}
 \label{tab:bulk}
 \begin{threeparttable}
  \begin{tabular}{lrrrrr}
   \toprule
   & Expt. & DFT & tabGAP-EAM & tabGAP & GAP-SOAP \\
   \midrule
    $a$ (Å) & 2.866\tnote{a} & 2.828\tnote{b} & 2.827 & 2.831 & 2.829 \\
    $B$ (GPa) & 169\tnote{a} & 195\tnote{b} & 240 & 193 & 195 \\
    $C_{11}$ (GPa) & 226\tnote{a}, 240\tnote{c} & 276\tnote{b} & 350 & 280 & 280 \\ 
    $C_{12}$ (GPa) & 140\tnote{a}, 136\tnote{c} & 155\tnote{b} & 186 & 150 & 153 \\ 
    $C_{44}$ (GPa) & 116\tnote{a}, 121\tnote{c} & 105\tnote{b} & 127 & 113 & 105  \\ 
    \\
    $E_\mathrm{surf}^{\hkl<100>}$ (meV/Å$^2$) & $\sim150$\tnote{d} & 162\tnote{b} & 140 & 160 & 163 \\
    $E_\mathrm{surf}^{\hkl<110>}$ (meV/Å$^2$) & $\sim150$\tnote{d} & 157\tnote{b} & 138 & 145 & 159 \\
    $E_\mathrm{surf}^{\hkl<111>}$ (meV/Å$^2$) & $\sim150$\tnote{d} & 175\tnote{b} & 162 & 173 & 175 \\
    \\
    $E^\mathrm{vac}_\mathrm{f}$ (eV) & $2.0 \pm 0.2$\tnote{e} & 2.29\tnote{b} & 2.12 & 2.23 & 2.24  \\
    $E^\mathrm{vac}_\mathrm{mig.}$ (eV) & $0.55 \pm 0.03$\tnote{f} & 0.68\tnote{g} & 0.66 & 0.72 & 0.64 \\
    $E^\mathrm{divac\text{-}1NN}_\mathrm{b}$ (eV) & & 0.16\tnote{h} & 0.07 & 0.13 & 0.13 \\
    $E^\mathrm{divac\text{-}2NN}_\mathrm{b}$ (eV) & & 0.23\tnote{h} & 0.33 & 0.23 & 0.23 \\
    $E^\mathrm{divac\text{-}3NN}_\mathrm{b}$ (eV) & & $-0.015$\tnote{h} & $-0.037$ & $-0.016$ & $-0.031$ \\
    $E^\mathrm{divac\text{-}4NN}_\mathrm{b}$ (eV) & & 0.05\tnote{h} & 0.03 & 0.03 & 0.02 \\
    $E^\mathrm{divac\text{-}5NN}_\mathrm{b}$ (eV) & & 0.06\tnote{h} & $-0.04$ & 0.02 & 0.05 \\
    \\
    $E^{\hkl<100>}_\mathrm{f}$ (eV) & & 5.46\tnote{i} & 5.15 & 5.27 & 5.23 \\
    $E^{\hkl<110>}_\mathrm{f}$ (eV) & & 4.32\tnote{i} & 4.30 & 4.26 & 4.06 \\
    $E^{\hkl<111>}_\mathrm{f}$ (eV) & & 5.09\tnote{i} & 4.77 & 5.06 & 4.94 \\
    $E^\mathrm{octa}_\mathrm{f}$ (eV) & & 5.56\tnote{i} & 5.11 & 5.41 & 5.31 \\
    $E^\mathrm{tetra}_\mathrm{f}$ (eV) & & 4.79\tnote{i} & 4.86 & 4.67 & 4.55 \\
    $E^\mathrm{SIA}_\mathrm{mig.}$ (eV) & $0.27 \pm 0.04$\tnote{f}, 0.32\tnote{f} & 0.34\tnote{i} & 0.29 & 0.30 & 0.31 \\
    \\
    $\alpha_L$ ($10^6$ K$^{-1}$) & 11.8\tnote{a} & & 10.5 & 11.3 & 11.6 \\
    $T_\mathrm{melt}$ (K) & 1811\tnote{a} & & 2020$\pm$20 & 1900$\pm$20 & 1900$\pm$20 \\
    $\Delta H_\mathrm{melt}$ (eV/atom) & 0.143\tnote{a} & & 0.18 & 0.25 & 0.23 \\
    $\rho_\mathrm{liq.}$ (at./Å$^3$) & 0.0759\tnote{j} & & 0.0780 & 0.0778 & 0.0765 \\
   \bottomrule
  \end{tabular}
  \begin{tablenotes}
   \item[a] Ref.~\cite{rumble_crc_2019}
   \item[b] This work.
   \item[c] Ref.~\cite{adams_elastic_2006}, 0 K data.
   \item[d] Ref.~\cite{tyson_surface_1977}
   \item[e] Ref.~\cite{de_schepper_positron_1983}
   \item[f] Ref.~\cite{takaki_resistivity_1983}
   \item[g] Ref.~\cite{ma_effect_2019}
   \item[h] VASP-PAW results from Ref.~\cite{malerba_comparison_2010}
   \item[i] Ref.~\cite{ma_universality_2019}
   \item[j] Ref.~\cite{assael_reference_2006}
  \end{tablenotes}
 \end{threeparttable}
\end{table*}

\begin{figure}
    \centering
    \includegraphics[width=\linewidth]{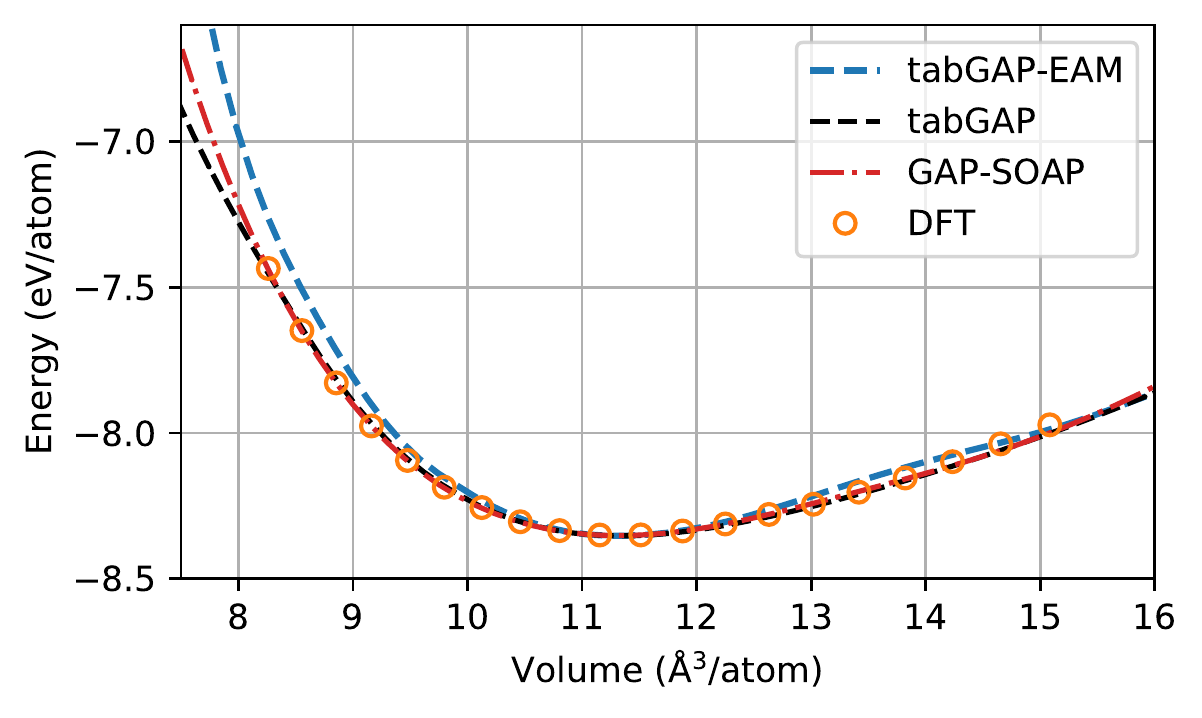}
    \caption{Energy versus volume per atom for ferromagnetic bcc iron compared between DFT and the three potentials.}
    \label{fig:ev}
\end{figure}

As a benchmark for how well the energy and force test errors translate to actual material properties, Tab.~\ref{tab:bulk} lists basic structural, elastic, surface, defect, and thermal properties of iron compared between experiment, DFT, and the three potentials (tabGAP-EAM, tabGAP, and GAP-SOAP). All three potentials reproduce the properties of iron well, with the noteworthy exception of the elastic constants by tabGAP-EAM. It overestimates all elastic constants by about 20\% compared to DFT, which already overestimates the experimental values by 10--20\%. This shortcoming of tabGAP-EAM is clear in the energy-volume curve of bcc iron shown in Fig.~\ref{fig:ev}, where the DFT points are well captured by tabGAP and GAP-SOAP, but tabGAP-EAM produces a stiffer curve around the equilibrium volume. Overall, from Tab.~\ref{tab:bulk} it is clear that tabGAP-EAM is by far the least accurate potential, as expected, while tabGAP and GAP-SOAP show in general very similar agreement with the reference data.

\begin{figure}
    \centering
    \includegraphics[width=\linewidth]{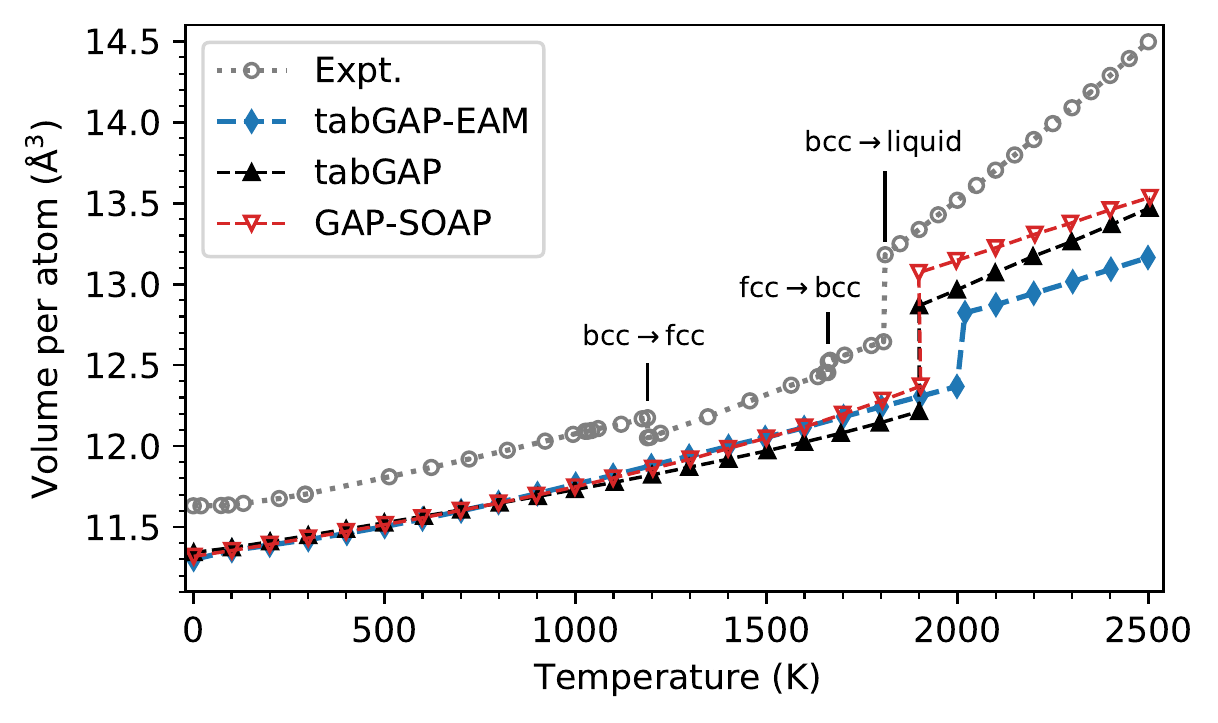}
    \caption{Thermal expansion of iron at zero pressure from the bcc phase to beyond melting. The phase transitions are indicated in the experimental curve. In the potentials, the bcc phase is stable until the melting temperature. The experimental data for the crystalline phases are from Ref.~\cite{basinski_lattice_1955} and combined with the liquid thermal expansion data from Ref.~\cite{assael_reference_2006}.}
    \label{fig:therm}
\end{figure}

On the other hand, the thermal expansion coefficient at 300 K as listed in Tab.~\ref{tab:bulk} is close to the experimental value in all three potentials. Furthermore, Fig.~\ref{fig:therm} shows the volume expansion at zero pressure for the temperature range from 0 K to far beyond the melting point for all three potentials and experimental measurements. All three potentials show very similar trends in the range of the ferromagnetic bcc phase. The experimental transition to the fcc phase and back to the bcc phase indicated in Fig.~\ref{fig:therm} is not captured by any of the tested potentials. With no magnetic degrees of freedom, this phase transition is not possible. However, the solid-liquid phase transition is captured the closest by the tabGAP and GAP-SOAP potentials, although the volume increase of the liquid phase above the melting point is slower with the temperature compared to the experiment, see Fig.~\ref{fig:therm}.

\subsection{Liquid properties}

The melting temperature predicted by both tabGAP and GAP-SOAP is 1900 K, only 5\% higher than the experimental 1811 K. The tabGAP-EAM potential overestimates it by 12\% (2020 K). The latent heat is also overestimated compared to experimental measurements (Tab.~\ref{tab:bulk}) in all three potentials, and is likely linked to the slight overestimation of the melting point~\cite{asadi_quantitative_2015}.

\begin{figure}
    \centering
    \includegraphics[width=\linewidth]{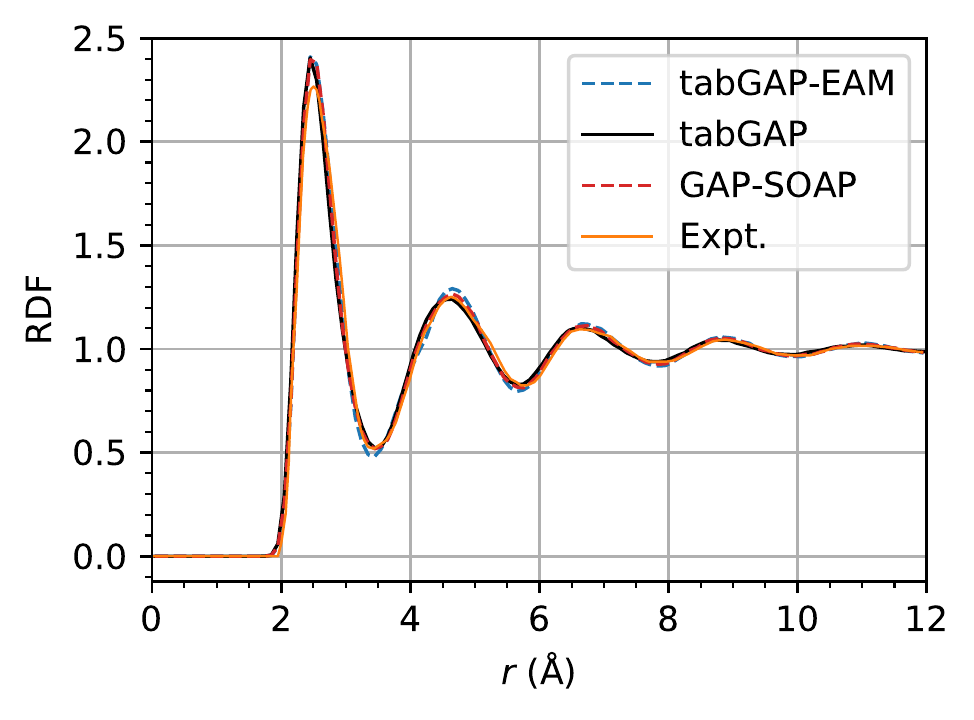}
    \caption{Radial distribution function (RDF) of the liquid phase, compared between the three potentials and experimental data obtained from Ref.~\cite{mendelev_development_2003}.}
    \label{fig:rdf}
\end{figure}

Fig.~\ref{fig:rdf} shows the radial distribution function computed as an average over time for an equilibrated liquid at the melting point in each potential. The tabGAP and GAP-SOAP data overlaps almost perfectly with experimental measurements~\cite{mendelev_development_2003}, and only the tabGAP-EAM potential shows small discrepancies.

\subsection{Repulsive potential}

\begin{figure*}
    \centering
    \includegraphics[width=0.34\linewidth]{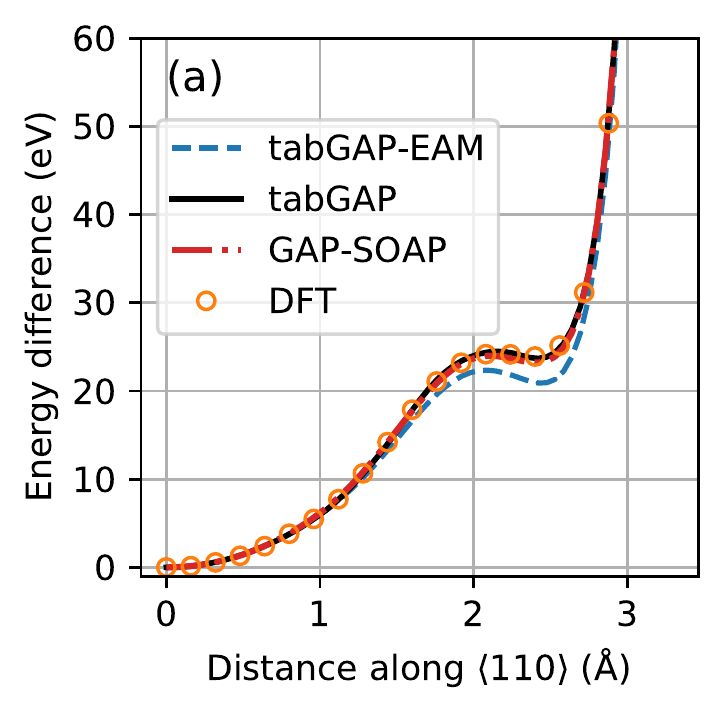}
    \includegraphics[width=0.65\linewidth]{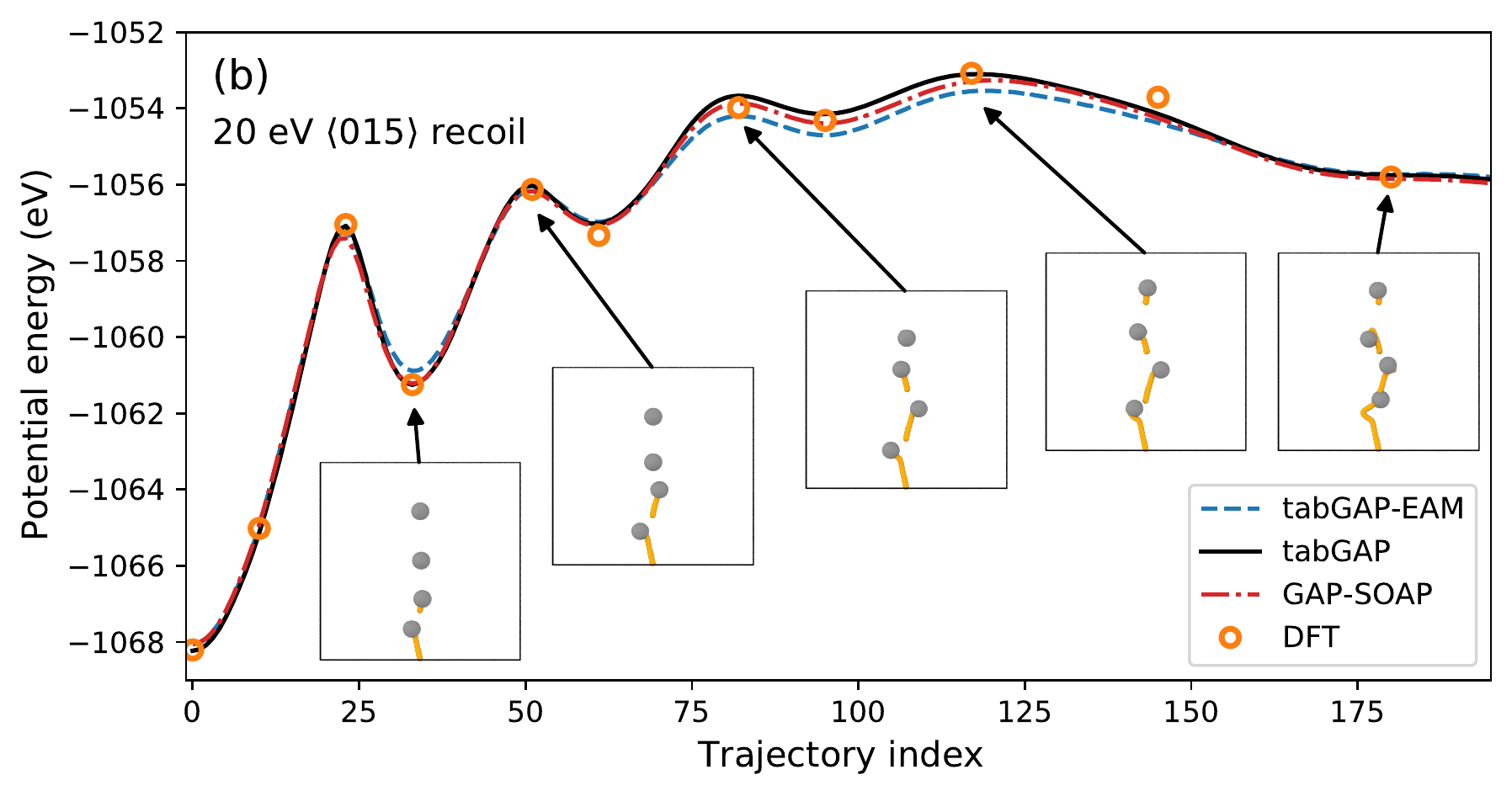}
    \caption{(a) Energy difference for stepwise movement of an atom in a rigid bcc lattice. (b) Energy landscape of a 20 eV \hkl<015> recoil in bcc iron, simulated with the tabGAP and compared with energies DFT and the other two potentials computed for the same trajectory. The insets visualise the evolution of the collision dynamics.}
    \label{fig:recoil}
\end{figure*}

We benchmarked the repulsive parts of the potentials both statically and dynamically. In the static test, an atom is moved step-wise along a given crystal direction while computing the change in energy. We chose to sample the \hkl<110> direction, as it provides an interesting energy landscape when the atom moves past its nearest neighbours. Reproducing the \hkl<110> energy landscape was also recently shown to correlate with other properties relevant for radiation damage simulations and was hence suggested as a good way to ensure that the repulsive part of the potential is accurate~\cite{becquart_modelling_2021}. Fig.~\ref{fig:recoil}(a) shows the results from all three potentials and DFT. Again, only tabGAP-EAM shows visible deviations from DFT while tabGAP and GAP-SOAP accurately follow the DFT points.

In the dynamic test, we simulated a low-energy recoil in a direction close to \hkl<100> with the tabGAP. The choice of direction, \hkl<015>, and recoil energy (20 eV) corresponds to a near-threshold event for defect creation (the minimum threshold displacement energy in Fe is around 20 eV and around the \hkl<100> direction~\cite{lomer_anisotropy_1967,maury_anisotropy_1976}). From the recoil simulation trajectory with the tabGAP, we recomputed the energies with the other two potentials and picked a set of interesting trajectory frames for DFT. The potential energy variation of the recoil trajectory is shown in Fig.~\ref{fig:recoil}(b), compared between the three potentials and DFT. All potentials are very close to the DFT points, suggesting that they can reliably model the interatomic collisions and initial defect creation processes that are important in collision cascade simulations.

\subsection{Defects}

Tab.~\ref{tab:bulk} lists basic properties of single vacancies, divacancies, and single self-interstitial atoms. GAP-SOAP and tabGAP compare well with the DFT data, in particular they reproduce accurate binding energies of divacancies and the correct order of stability and energy differences of single SIA configurations.

\begin{figure}
    \centering
    \includegraphics[width=\linewidth]{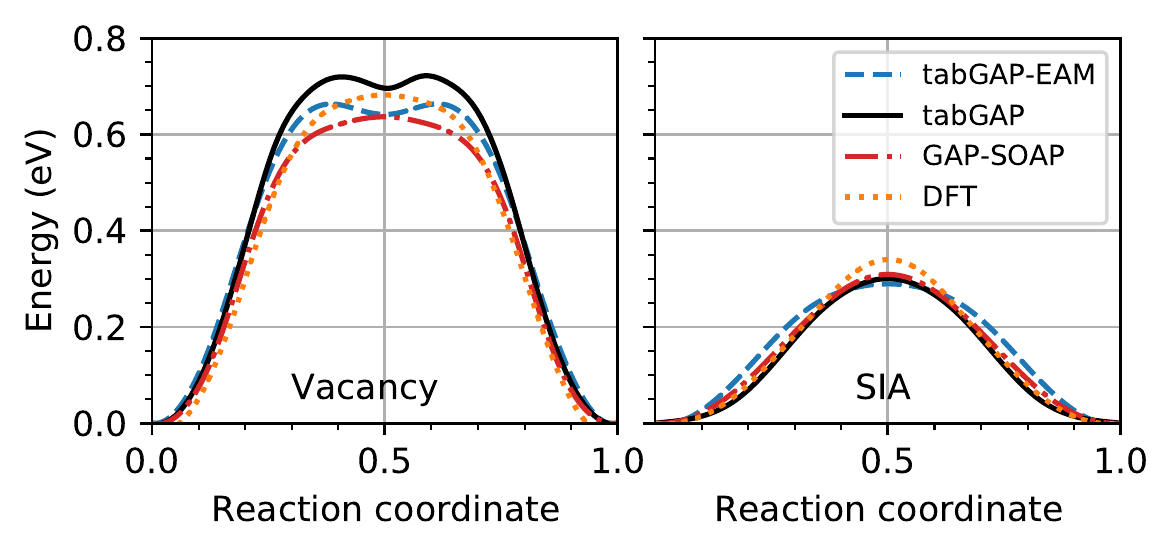}
    \caption{Barrier for single vacancy and self-interstitial atom (SIA) migration in bcc iron. The DFT data are from Refs.~\cite{ma_effect_2019,ma_universality_2019}.}
    \label{fig:neb}
\end{figure}

The migration barriers of the single vacancy and SIA, computed with the NEB method, are shown in Fig.~\ref{fig:neb} compared to DFT data~\cite{ma_effect_2019,ma_universality_2019}. All three potentials reproduce the migration energies well, although tabGAP-EAM and tabGAP show a double-hump profile for the vacancy migration barrier which is not present in DFT but is a common feature of existing interatomic potentials~\cite{malerba_comparison_2010}.

\begin{figure}
    \centering
    \includegraphics[width=\linewidth]{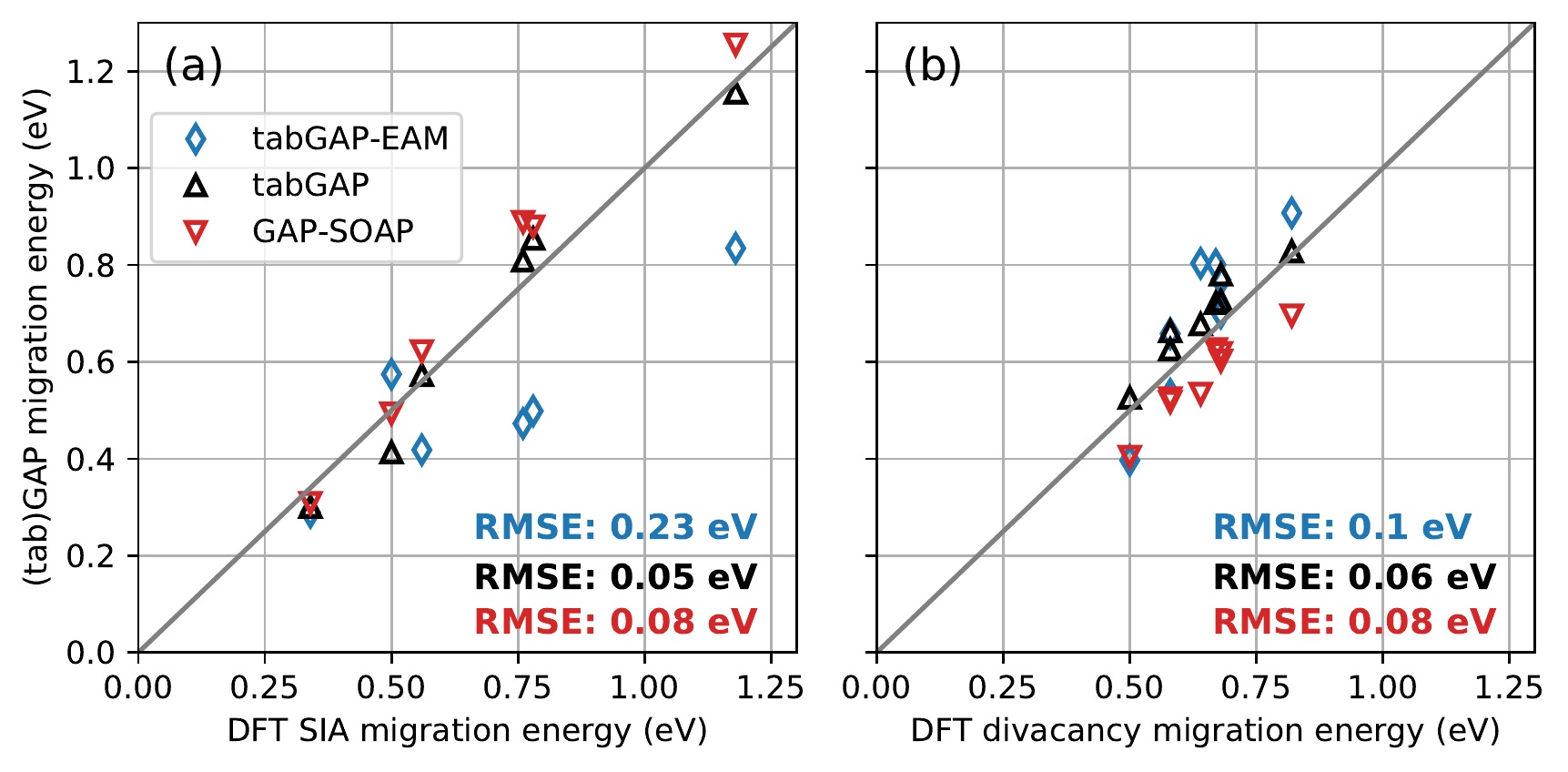}
    \caption{Comparison between DFT data from Ref.~\cite{malerba_ab_2010} and NEB calculations with the three potentials for various migration barriers of (a): a single \hkl<110> SIA, and (b): divacancies. For illustrations of the various migration paths as well as all migration barriers and energies, see the Supplemental material online.}
    \label{fig:neb_all}
\end{figure}

Fig.~\ref{fig:neb_all} compares various migration barriers of single SIAs and divacancies between the potentials and DFT data from Ref.~\cite{malerba_comparison_2010}. The migration paths and corresponding energies are illustrated and listed in the Supplemental material. Fig.~\ref{fig:neb_all} shows that overall, tabGAP and GAP-SOAP produce migration energies that are quite consistent with the DFT data. GAP-SOAP has a tendency to slightly underestimate (di)vacancy migration energies and shows a RMS error of 0.08 eV compared to DFT for both SIAs and divacancies. The tabGAP is somewhat more accurate with RMS errors 0.05--0.06 eV, while tabGAP-EAM performs reasonably well for divacancy migration but quite poorly for SIA migration paths.

\begin{table*}
 \centering
 \caption{Formation energies of the most stable small parallel and nonparallel SIA clusters in iron. The subscripts indicate the number of SIAs in the cluster. The formation energies of the nonparallel clusters are shown as differences to the parallel \hkl<110> configurations, so that negative values indicates more stable than the parallel cluster. For illustrations of the nonparallel configurations, see e.g. Ref.~\cite{dezerald_stability_2014}.}
 \label{tab:multi_SIA}
  \begin{tabular}{lrrrr}
   \toprule
   & DFT~\cite{malerba_comparison_2010,dezerald_stability_2014} & tabGAP-EAM & tabGAP & GAP-SOAP \\
   \midrule
    $\hkl<110>_2$ parallel & 6.99--7.55 & 7.86 & 7.80 & 7.37 \\
    $\hkl<110>_2$ triangle & $-0.1$ & 0.18 & $-0.17$ & $-0.15$  \\
    C15$_2$ & 0.8 & 1.68 & 1.19 & 0.75  \\
   \midrule
    $\hkl<110>_3$ parallel & 9.89--10.39 & 10.91 & 11.12 & 10.44 \\
    $\hkl<110>_3$ hexagon & $-0.06$ & 0.78 & $-0.04$ & 0.05 \\
   \midrule
    $\hkl<110>_4$ parallel & 12.31--13.60 & 13.62 & 14.05 & 13.32 \\
    C15$_4$ & $-1.29$, $-1.83$ & $-0.14$ & $-1.58$ & $-1.78$ \\
   \bottomrule
  \end{tabular}
\end{table*}

\begin{figure}
    \centering
    \includegraphics[width=\linewidth]{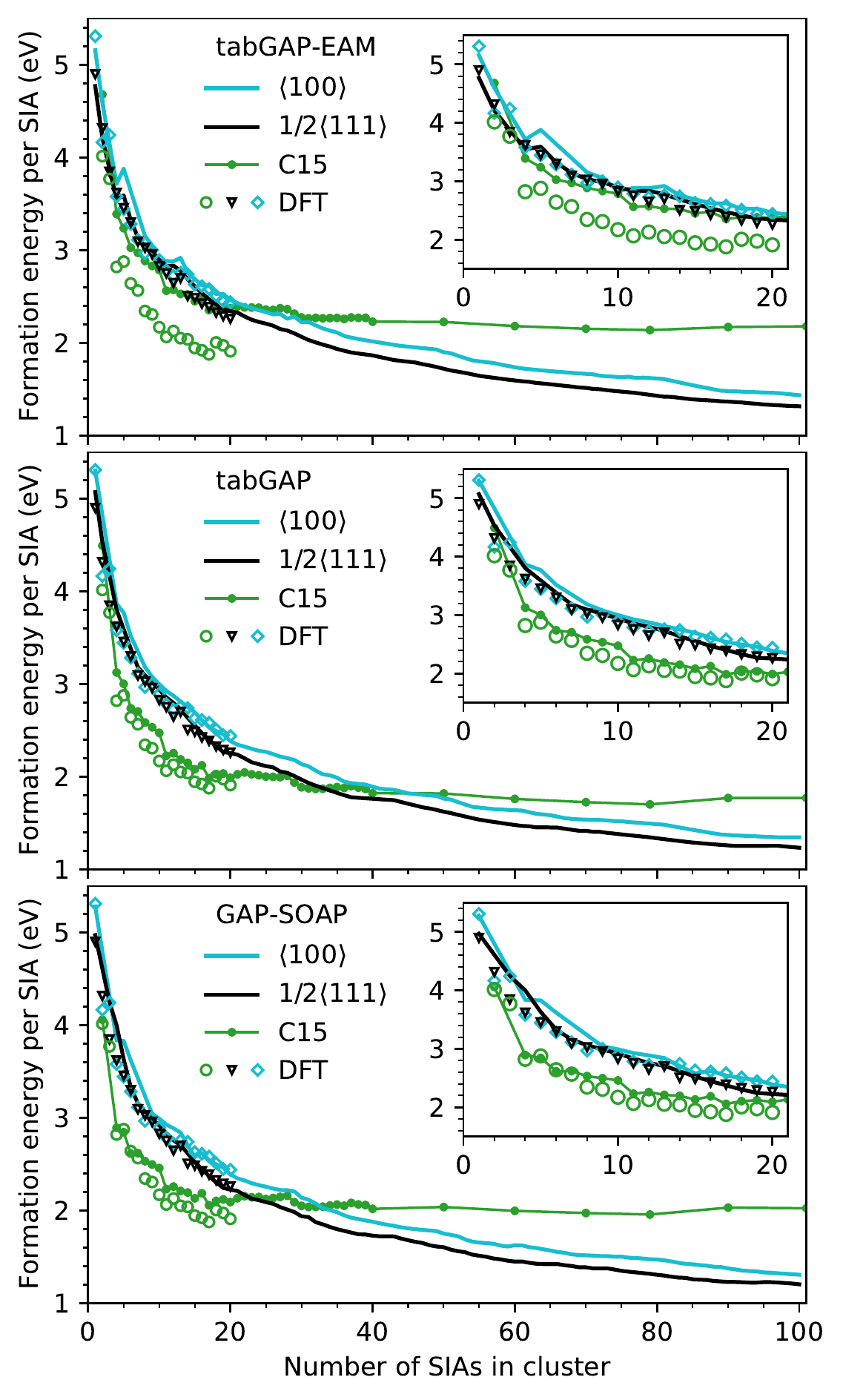}
    \caption{Formation energy per atom of SIA clusters in iron (C15 clusters, 1/2\hkl<111>, and \hkl<100> dislocation loops). The three potentials are compared with DFT data from Ref.~\cite{alexander_ab_2016}. The insets show zoomed-in views of the DFT-computed size range.}
    \label{fig:clust}
\end{figure}

The energy landscape and possible geometries of self-interstitial clusters in iron is rich and challenging for classical interatomic potentials to reproduce. It is now well established by DFT calculations that nonparallel clusters are at small sizes much more energetically stable than parallel dumbbells and dislocation loops~\cite{marinica_irradiation-induced_2012,dezerald_stability_2014,alexander_ab_2016}, in contrast to nonmagnetic bcc metals like W. For clusters of 2 and 3 SIAs, the triangular and hexagonal configurations of \hkl<110> dumbbells lying in a \hkl{111} plane are the most stable SIA clusters~\cite{dezerald_stability_2014}. For sizes 4 and larger, SIA clusters with the C15 Laves crystal symmetry, made up of these triangular and hexagonal building blocks, become the most stable. At large sizes, 1/2\hkl<111> dislocation loops become the most stable SIA cluster, like in all other bcc metals. There have been several attempts to reproduce this complex landscape of SIA clusters in iron with analytical interatomic potentials~\cite{marinica_irradiation-induced_2012,byggmastar_dynamical_2020,alexander_interatomic_2020}, although none have been completely successful. GAP-SOAP and tabGAP provide improvements over the existing analytical potentials, but still leaves some room for improvement.

Table~\ref{tab:multi_SIA} lists formation energies of parallel and the nonparallel \hkl<110> dumbbell configurations discussed above. Both tabGAP and GAP-SOAP correctly reproduce the triangular configuration as the most stable cluster of 2 SIAs. Only tabGAP predicts the hexagonal size-3 cluster to be the most stable, although the energy difference compared to the parallel configuration is very small also in GAP-SOAP. For size 4, both tabGAP and GAP-SOAP correctly reproduce the C15 clusters to be significantly more stable than parallel dumbbells. From Tab.~\ref{tab:multi_SIA} it is clear that tabGAP-EAM, due to its limited flexibility and lack of angular dependence, struggles to reproduce the relative formation energies of SIA clusters.

Most of the small SIA clusters discussed above are well-covered by the training database and good accuracy is therefore to be expected. We also computed the formation energies of clusters up to 100 SIAs in the form of C15 clusters and dislocation loops with the \hkl<100> and 1/2\hkl<111> Burgers vectors. Fig.~\ref{fig:clust} shows the formation energies per interstitial in all three potentials. The results are compared to DFT data for small clusters from Ref.~\cite{alexander_ab_2016}. Note that there are often many geometrically different ways to construct clusters of a given size. Hence, our clusters may not be exactly the same as the DFT data used for comparison~\cite{alexander_ab_2016}. For dislocation loops at sizes larger than a few interstitials, the difference in formation energy between different configuration is typically quite small. For C15 clusters, however, there are vast amounts of possible configurations for any given size and the formation energy may vary significantly. Only a few sizes allow for well-defined high-symmetry shapes, which have relatively low formation energy. Previous work, including the DFT work with which we compare with here, have employed various criteria for constructing possible low-energy C15 clusters. Here, we use a growth-annealing method as described in section~\ref{sec:md} to find low-energy C15 clusters. We only report the formation energy of the lowest-energy C15 cluster that we found at each size in Fig.~\ref{fig:clust} (which, however, is most likely not the most stable out of all theoretical C15 configurations and also likely not the same configuration as in the DFT study).

Fig.~\ref{fig:clust} shows that the tabGAP and GAP-SOAP reproduce the relative stability of the three clusters in good agreement with the DFT data. The formation energies of C15 clusters are somewhat overestimated, and very much so in tabGAP-EAM, which does not provide any improvement over existing EAM potentials~\cite{marinica_irradiation-induced_2012,alexander_interatomic_2020}. 1/2\hkl<111> loops are more stable than \hkl<100> loops for the entire size range in all potentials, consistent with DFT extrapolation~\cite{alexander_ab_2016}. The DFT-based extrapolation model developed in Ref.~\cite{alexander_ab_2016} suggests a crossover in the energy of C15 and 1/2\hkl<111> loops at around 50 interstitials and between C15 and \hkl<100> loops at around 90 interstitials. The tabGAP and GAP-SOAP predict the corresponding crossovers at much lower sizes, 34 and around 45 for tabGAP, and 23 and 33 for GAP-SOAP. In comparison, the recent ML potentials based in linear regression achieved a crossover between C15 and 1/2\hkl<111> at around 40 SIAs~\cite{goryaeva_efficient_2021}, i.e. somewhat closer to the DFT estimate than tabGAP. Given that the stability of C15 and other clusters can vary significantly between different exchange-correlation functionals and pseudo- or PAW potentials in DFT~\cite{dezerald_stability_2014}, it remains unclear if the differences in crossovers is a shortcoming of the potentials, or if the difference can to some extent be attributed to differences in our DFT compared to the reference DFT data from Ref.~\cite{alexander_ab_2016}.

It is noteworthy that while GAP-SOAP is most accurate among the potentials for small clusters (Tab.~\ref{tab:multi_SIA}), the tabGAP shows better transferability to larger clusters (Fig.~\ref{fig:clust}).

\subsection{Screw dislocations}

\begin{figure}
    \centering
    \includegraphics[width=0.45\linewidth]{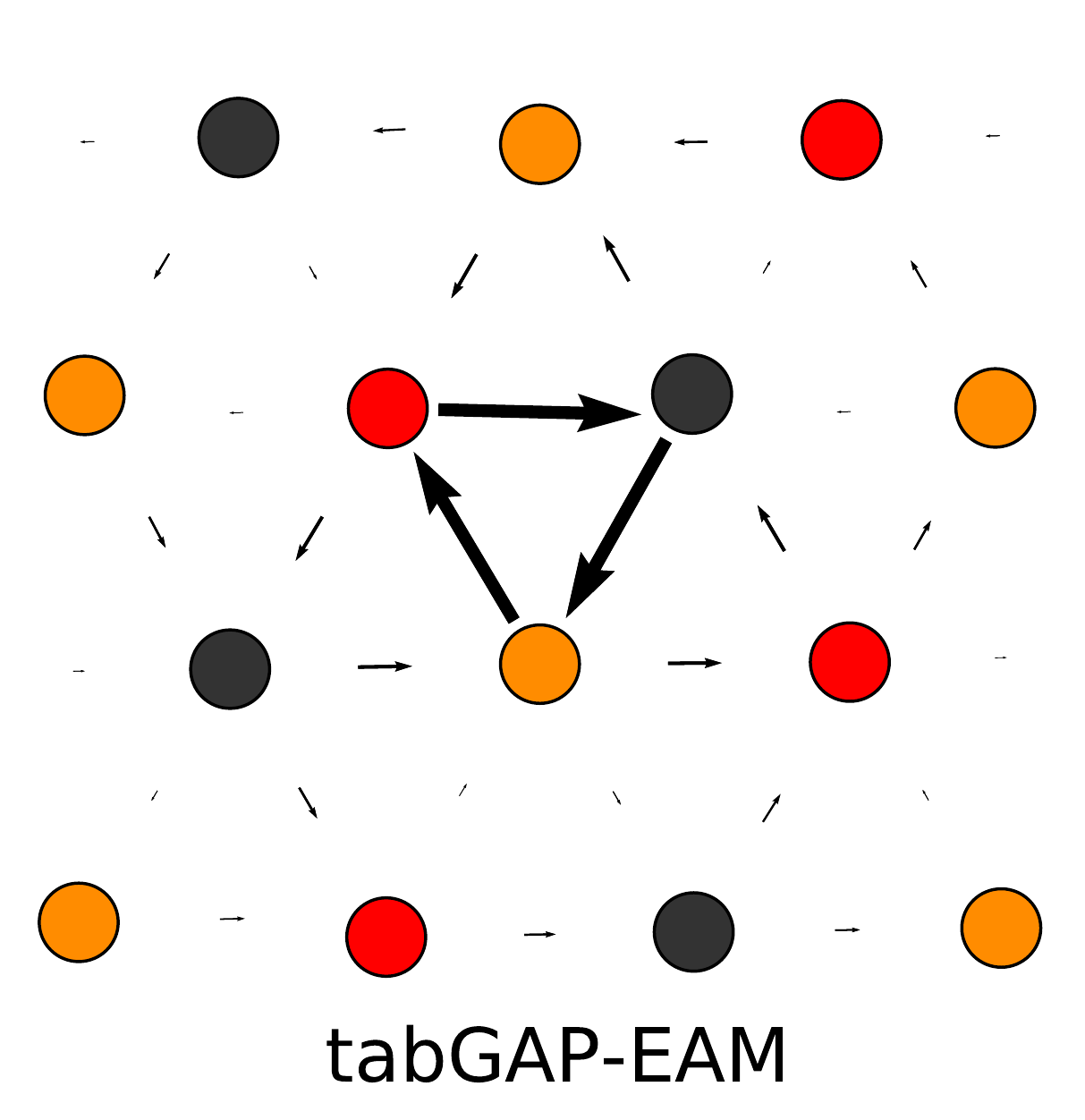}
    \hfill
    \includegraphics[width=0.45\linewidth]{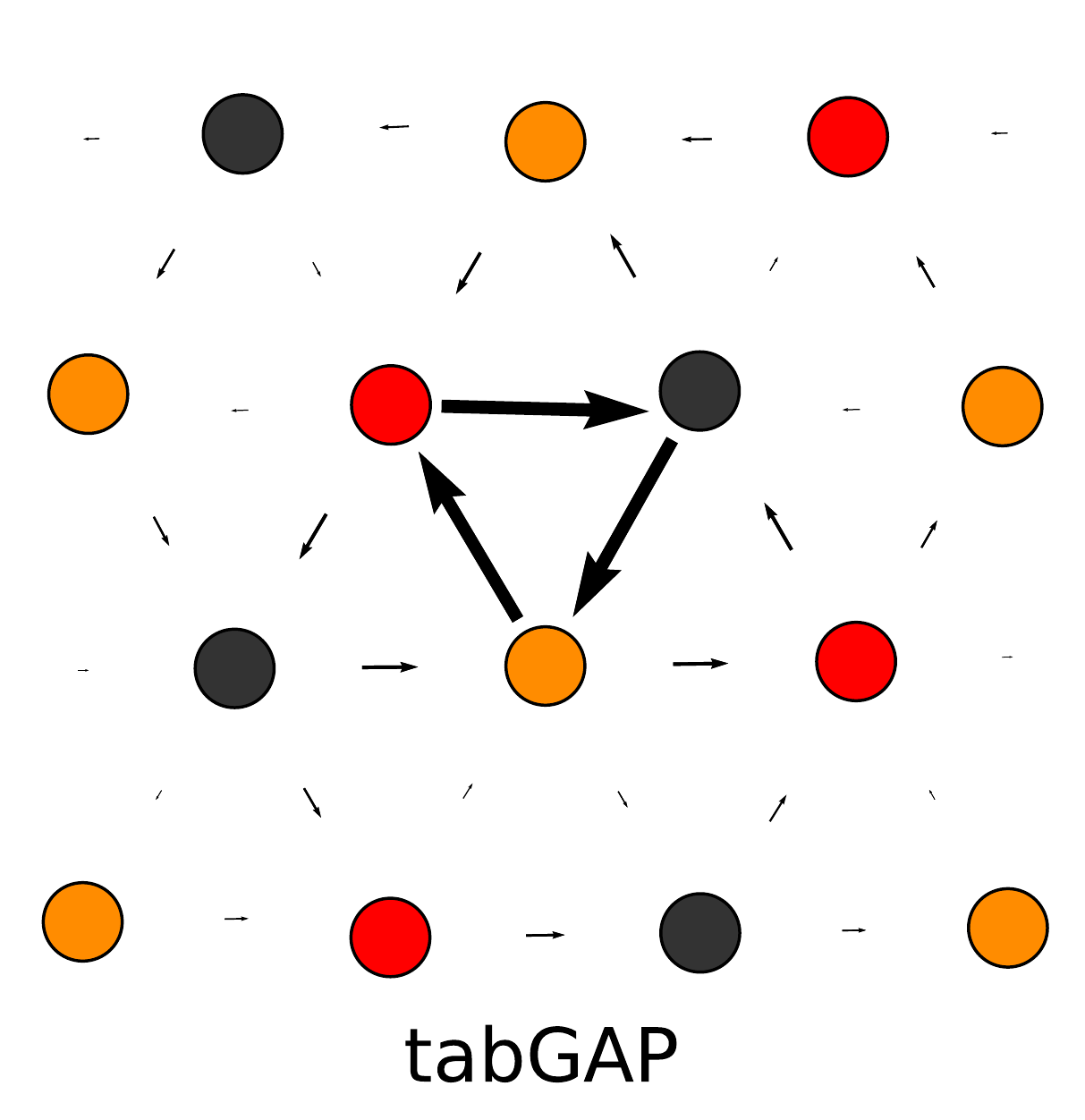}
    \includegraphics[width=0.45\linewidth]{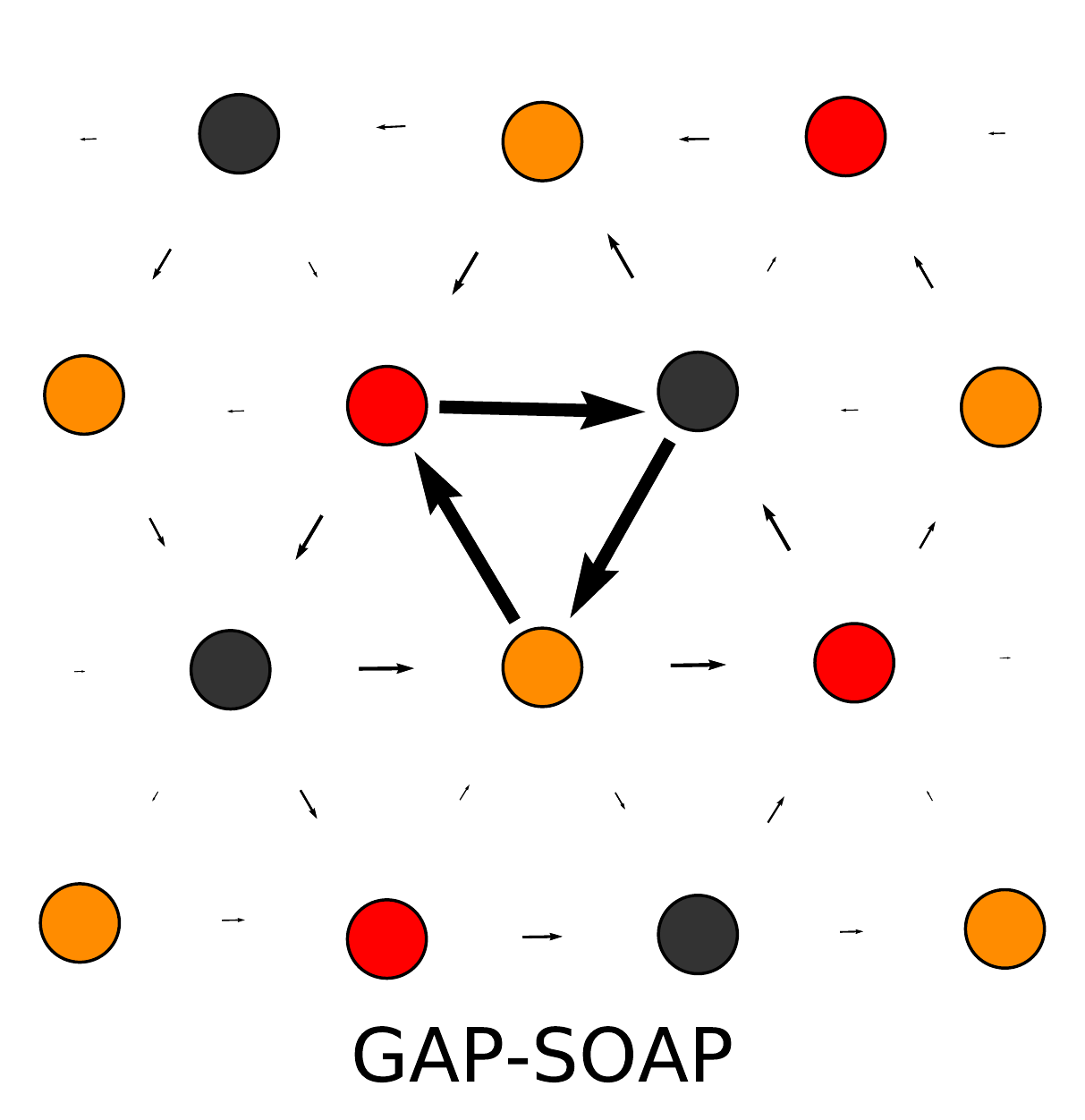}
    \hfill
    \includegraphics[width=0.45\linewidth]{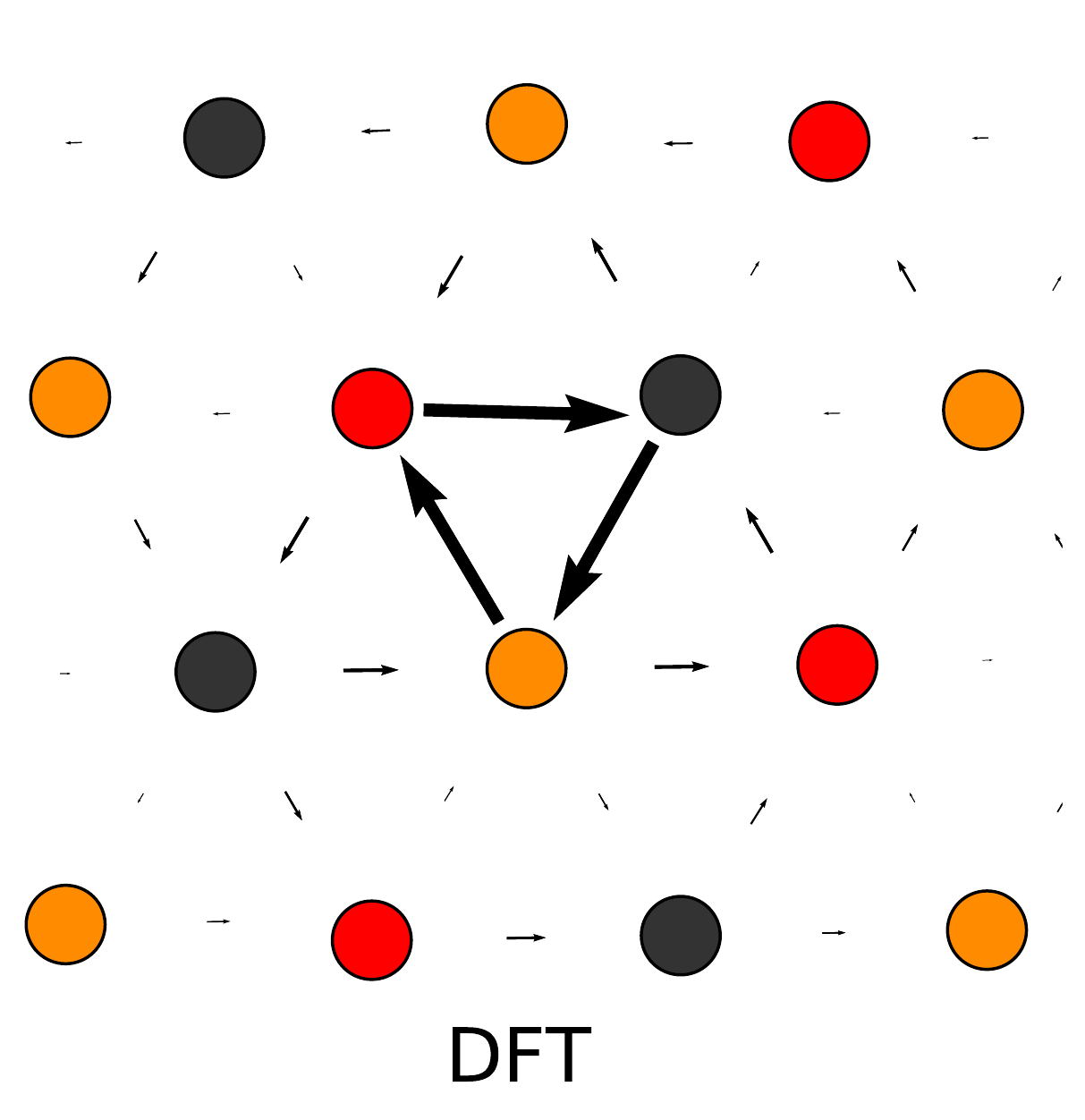}
    \caption{Relaxed core structure of a 1/2\hkl<111> screw dislocation in the three potentials compared to DFT, visualised as the commonly used differential displacement plots~\cite{vitek_core_1970}. The colours indicate the three different \hkl(111) layers spanning one Burgers vector length. The arrows are drawn between nearest neighbours and indicate the out-of-plane \hkl<111> displacements with respect to the perfect bulk.}
    \label{fig:core}
\end{figure}

\begin{figure}
    \centering
    \includegraphics[width=\linewidth]{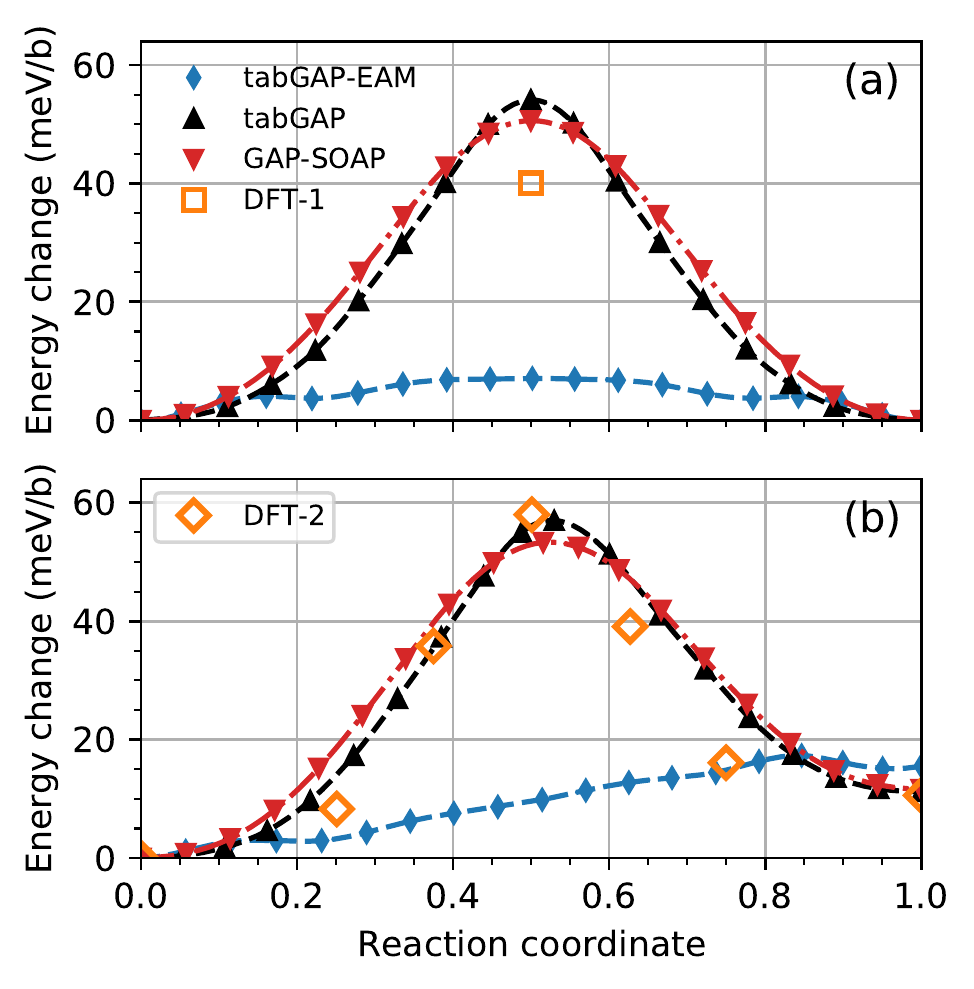}
    \caption{Peierls barrier for 1/2\hkl<111> screw dislocation migration obtained from NEB calculations of a migrating dislocation dipole. In (a), the dislocations migrate simultaneously while in (b), they migrate one by one. The barriers are compared with DFT data obtained in the same ways (DFT-1: \cite{ventelon_ab_2013}, DFT-2: \cite{dragoni_achieving_2018}).}
    \label{fig:peierls}
\end{figure}

Reproducing the basic properties of screw dislocations is often challenging for traditional interatomic potentials. We confirmed that the tabGAPs and GAP-SOAP all reproduce the symmetric nondegenerate core structure of the 1/2\hkl<111> screw dislocation as predicted by DFT. Fig~\ref{fig:core} shows the relaxed core of the screw dislocation in all three potentials and our DFT. We used 135-atom boxes with the quadrupolar periodic arrangement of screw dislocation dipoles~\cite{ventelon_ab_2013}, produced by inserting two screw dislocations (around 17 Å apart) with opposite Burgers vectors.

We also computed the Peierls barrier for screw dislocation migration in the tabGAPs and GAP-SOAP using the NEB method. Fig.~\ref{fig:peierls} shows the results. The barriers are computed in two ways, with simultaneous migration of both dislocations (Fig.~\ref{fig:peierls}a), and with only one of the dislocations migrating (Fig.~\ref{fig:peierls}b). The latter approach replicates the method used in Ref.~\cite{dragoni_achieving_2018}, which allows a direct comparison between the potentials and their DFT barrier. We used the same 135-atom box to be consistent with the DFT results. The obtained Peierls barriers from simultaneous migration (Fig.~\ref{fig:peierls}a) are compared to DFT data from Ref.~\cite{ventelon_ab_2013}.

Fig.~\ref{fig:peierls} shows that both tabGAP and GAP-SOAP produce similar barriers with shapes and heights consistent with the DFT results. The tabGAP-EAM potential, like most existing EAM potentials, fails to reproduce the Peierls barrier and predicts an almost flat energy barrier. The tabGAP and GAP-SOAP agrees much better with the DFT barrier from Ref.~\cite{dragoni_achieving_2018}, which we believe is more consistent with our DFT training data.

\section{Summary and outlook}

We have developed four interatomic potentials for iron using machine-learning methods and increasingly flexible combinations of descriptors for the local atomic environments. Three out of these potentials were thoroughly benchmarked, and two of them (tabGAP and GAP-SOAP) showed overall great accuracy for a range of solid and liquid properties. The three tested potentials span more than three orders of magnitude in computational cost, and hence provide options depending on the required accuracy and speed. All potentials contain accurate repulsive parts that make them applicable to collision cascade simulations.

The results demonstrate that our method for tabulation of low-dimensional Gaussian approximation potentials (tabGAP) provide interatomic potentials with a good balance between accuracy, speed, and transferability. The tabGAP combines simple two-body, three-body, and EAM-like density descriptors that together provide good flexibility and can be mapped onto suitable grids, making them computationally efficient. In particular, we showed that our new simple EAM-like descriptor provides the many-body coordination dependence necessary to accurately describe the liquid phase. The tabGAP developed here shows overall similar accuracy to the GAP-SOAP potential but at a much lower computational cost, similar to that of classical analytical angular-dependent potentials. Given its modest computational cost and good accuracy and transferability for defect properties, the tabGAP is well-suited for large-scale radiation damage simulations.

\section*{Acknowledgments}

This work has received funding from the Academy of Finland through the HEADFORE project (grant number 1333225).
The authors wish to thank the Finnish Computing Competence Infrastructure (FCCI) and CSC -- IT Center for Science for supporting this project with computational and data storage resources.
This work has been carried out within the framework of the EUROfusion Consortium, funded by the European Union via the Euratom Research and Training Programme (Grant Agreement No 101052200 — EUROfusion). Views and opinions expressed are however those of the author(s) only and do not necessarily reflect those of the European Union or the European Commission. Neither the European Union nor the European Commission can be held responsible for them.

\appendix

\section{Machine-learned EAM functions}

\begin{figure}
    \centering
    \includegraphics[width=\linewidth]{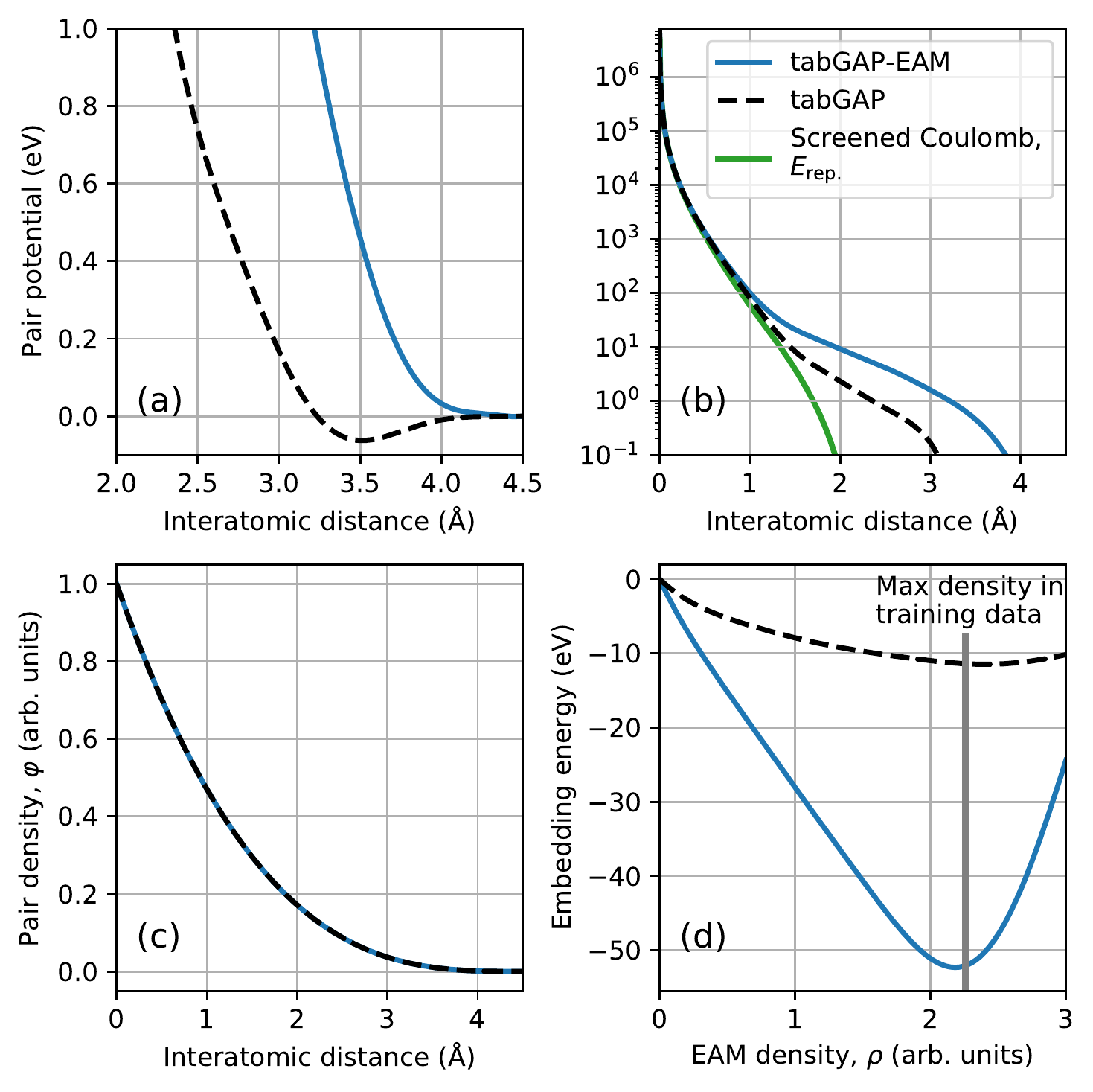}
    \caption{The machine-learned pair potentials (a-b), the pairwise density function used in the descriptor (c), and the machine-learned embedding functions of the EAM terms in the tabGAP-EAM and tabGAP. The vertical grey line in (d) is the maximum total density present in the local atomic environments of the training structures.}
    \label{fig:mleam}
\end{figure}

Fig.~\ref{fig:mleam} shows the machine-learned EAM functions of the tabGAP-EAM and tabGAP (note that the latter also contains the separate three-body term). The pair potential is machine-learned and connects smoothly to the repulsive screened Coulomb potential. The pair density function is fixed as part of the descriptor, as described in the Methods section, and is identical for both potentials. The machine-learned embedding functions shown in Fig.~\ref{fig:mleam}(d) have the physically reasonable monotonically decreasing but convex shape~\cite{mendelev_development_2003}. The grey vertical line in Fig.~\ref{fig:mleam}(d) indicates the maximum total density encountered in the training structures, after which the embedding energy starts approaching zero due to lack of training points (densities higher than this will in practice never be encountered as it would require multiple atoms simultaneously very close to each other).

\section{tabGAP grid convergence}

\begin{figure}
    \centering
    \includegraphics[width=\linewidth]{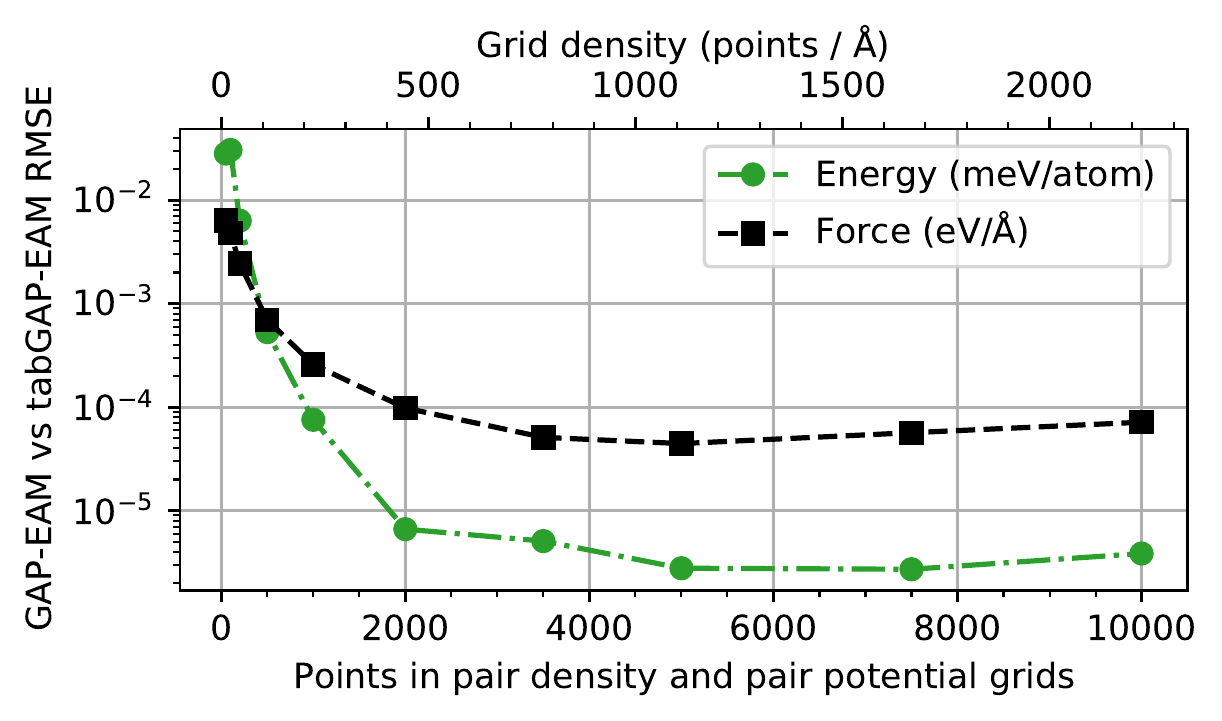}
    \includegraphics[width=\linewidth]{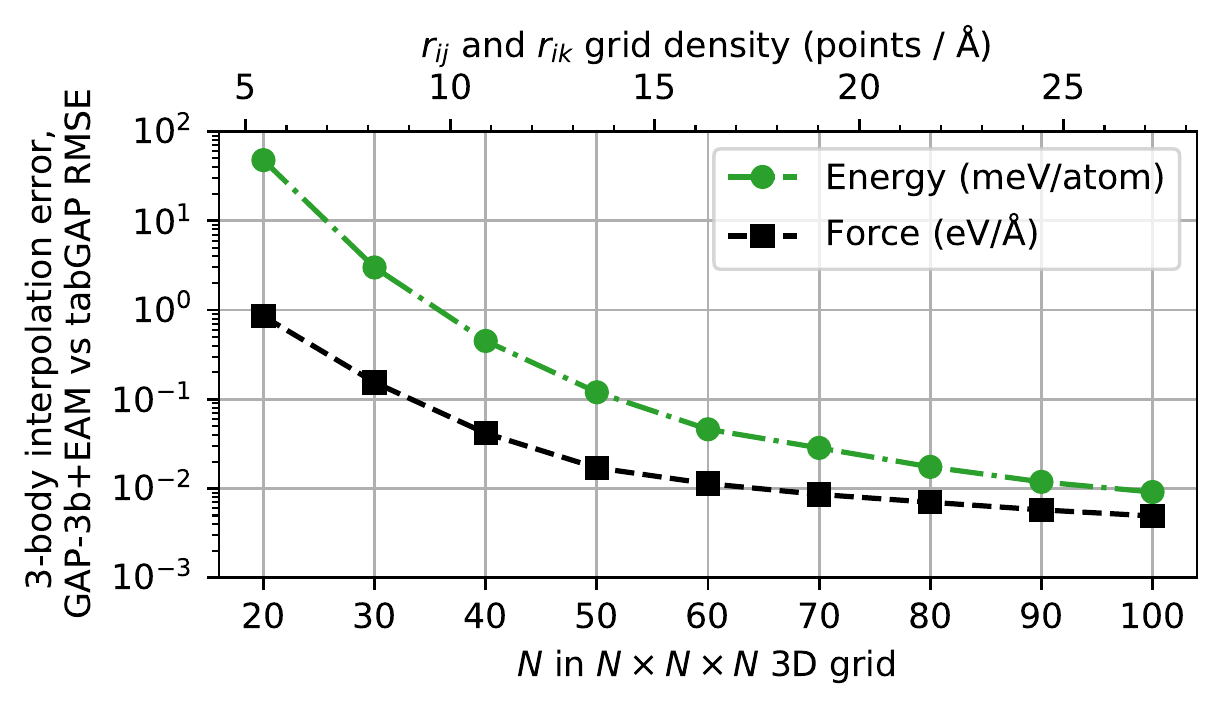}
    \caption{Convergence of the cubic-spline interpolation errors for the 1D functions (pair potential and embedding energy) and the 3D three-body term. For the final tabGAPs, we used 5000 points for the 1D interpolation and $80\times 80 \times 80$ points for the 3D grid.}
    \label{fig:grid}
\end{figure}

Fig.~\ref{fig:grid} shows the convergence of the tabGAP interpolation error as functions of grid size. For 1D interpolation, the errors are already vanishingly small when using more than a few hundred points. For the final tabGAPs, we used 5000 points. For the 3D $(r_{ij}, r_{jk}, \cos \theta_{ijk})$ grid, using thousands of points in each dimension is out of reach, but Fig.~\ref{fig:grid} shows that the interpolation error is already negligible compared to the accuracy of the potential itself when using more than 50 grid points in each dimension. For the final tabGAP, we used a $N\times N \times N$ grid with $N=80$, for which the interpolation error is well below 0.1 meV/atom and 0.01 eV/Å.

\bibliography{mybib}

\end{document}


\title{Supplemental material for: Multiscale machine-learning interatomic potentials for ferromagnetic and liquid iron}

\author{J. Byggmästar}
\thanks{Corresponding author}
\email{jesper.byggmastar@helsinki.fi}
\affiliation{Department of Physics, P.O. Box 43, FI-00014 University of Helsinki, Finland}
\author{G. Nikoulis}
\affiliation{Department of Physics, P.O. Box 43, FI-00014 University of Helsinki, Finland}
\author{A. Fellman}
\affiliation{Department of Physics, P.O. Box 43, FI-00014 University of Helsinki, Finland}
\author{F. Granberg}
\affiliation{Department of Physics, P.O. Box 43, FI-00014 University of Helsinki, Finland}
\author{F. Djurabekova}
\affiliation{Department of Physics, P.O. Box 43, FI-00014 University of Helsinki, Finland}
\affiliation{Helsinki Institute of Physics, Helsinki, Finland}
\author{K. Nordlund}
\affiliation{Department of Physics, P.O. Box 43, FI-00014 University of Helsinki, Finland}

\date{\today}

\maketitle

\section{Migration barriers}

\begin{figure}
    \centering
    \begin{subfigure}[h]{0.32\linewidth}
     \centering
     \includegraphics[width=\textwidth]{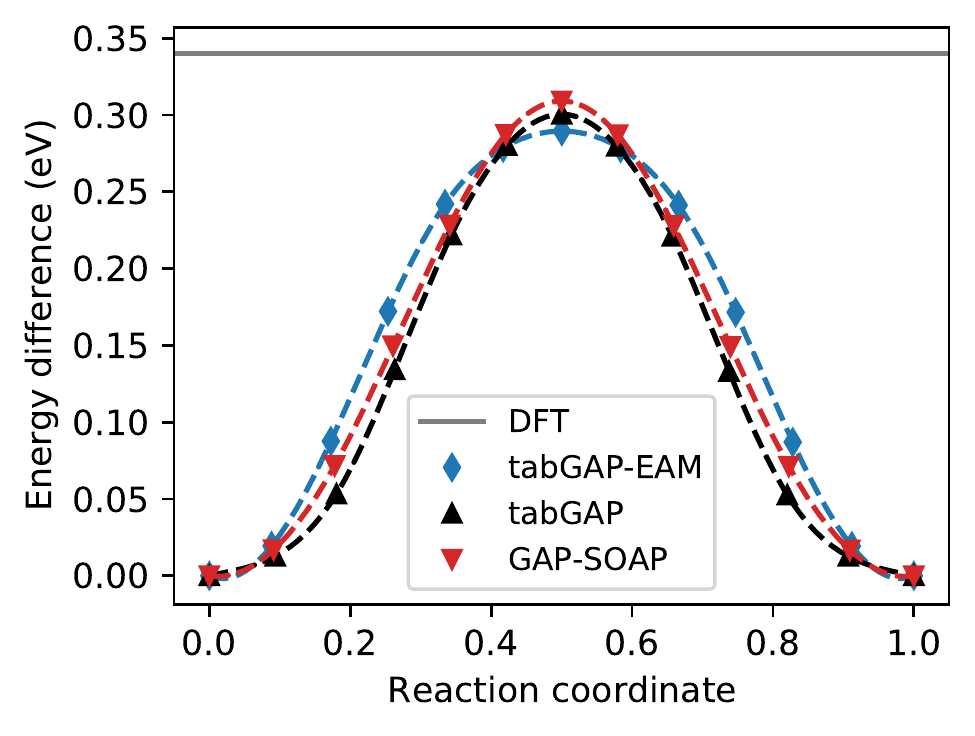}
     \includegraphics[width=0.9\textwidth]{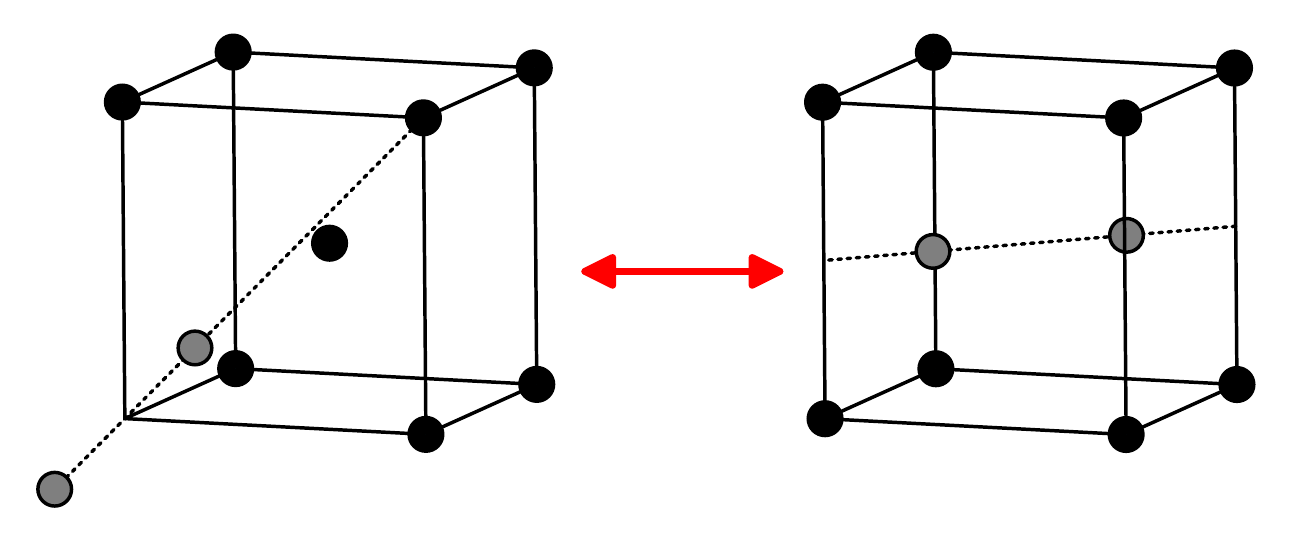}
     \caption{Main path: 1NN \hkl[110] $\rightarrow$ \hkl[101] jump.}
    \end{subfigure}
    \begin{subfigure}[h]{0.32\linewidth}
     \centering
     \includegraphics[width=\textwidth]{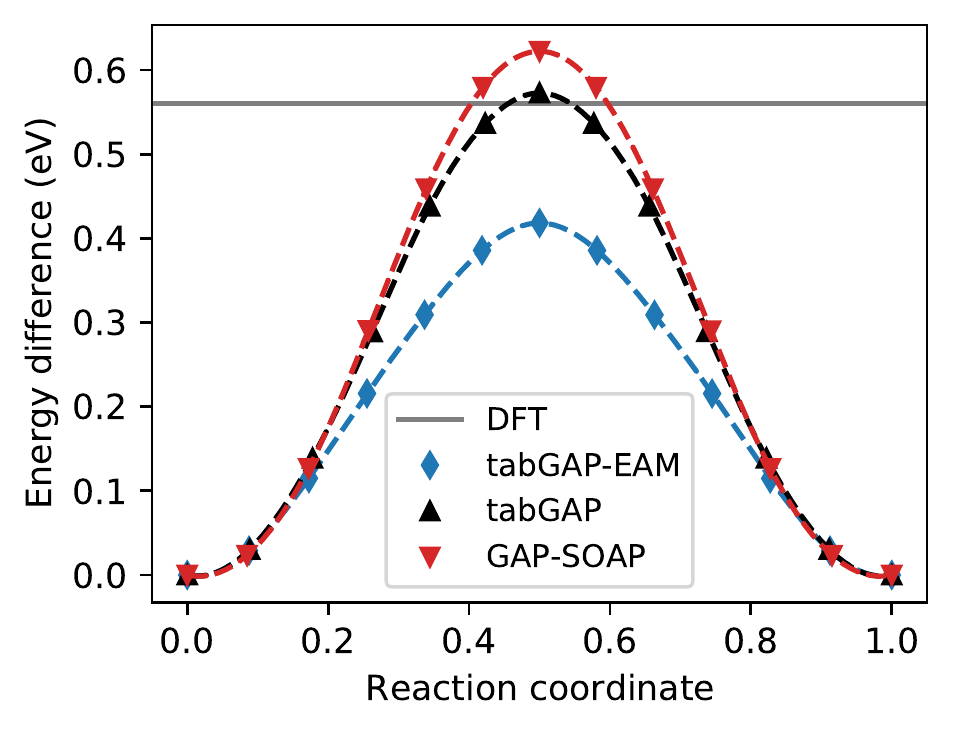}
     \includegraphics[width=0.9\textwidth]{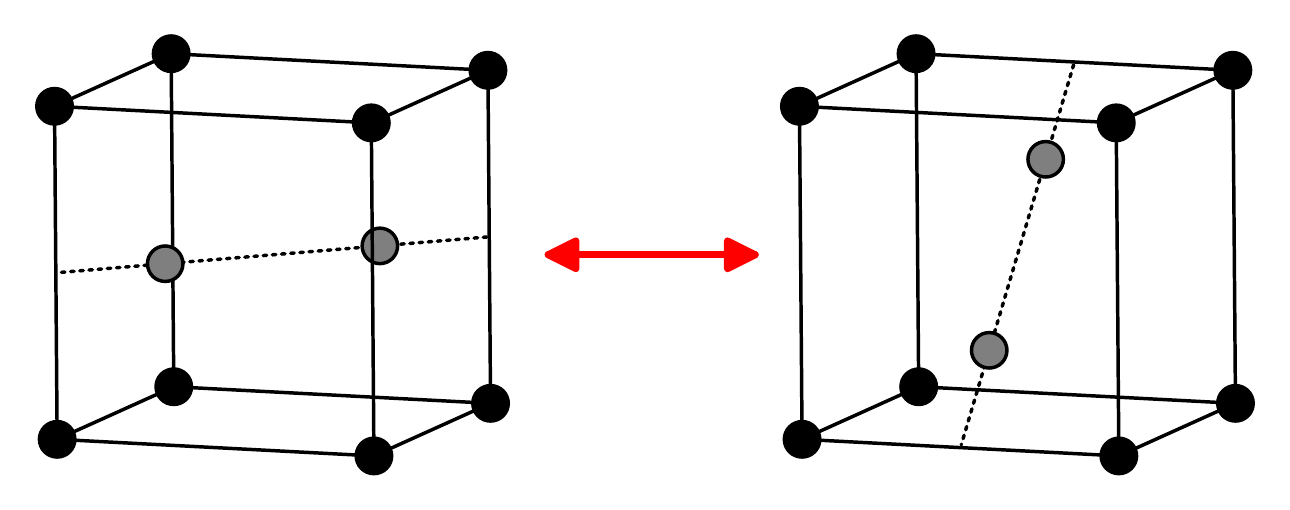}
     \caption{On-site rotation \hkl[101] $\rightarrow$ \hkl[011].}
    \end{subfigure}
    \begin{subfigure}[h]{0.32\linewidth}
     \centering
     \includegraphics[width=\textwidth]{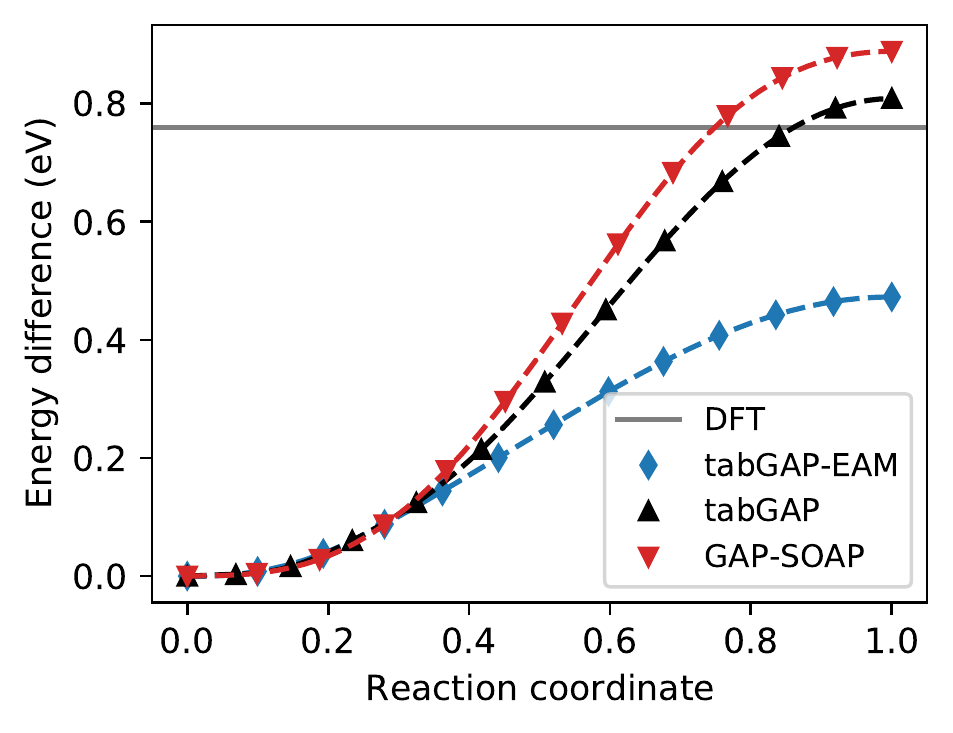}
     \includegraphics[width=0.9\textwidth]{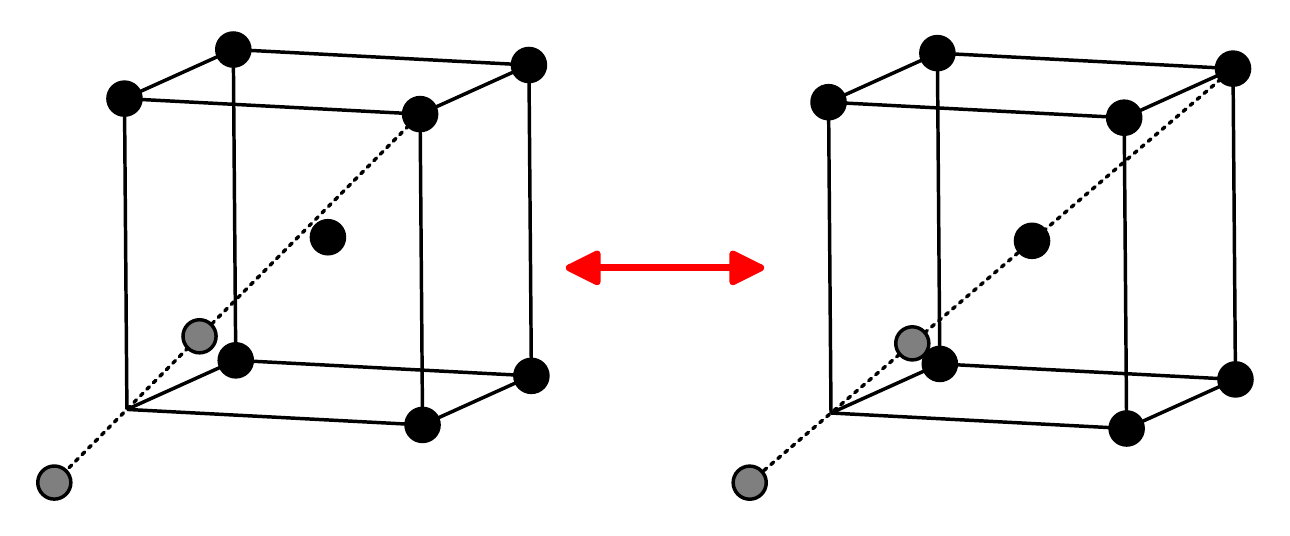}
     \caption{On-site rotation \hkl[110] $\rightarrow$ \hkl[111].}
     \end{subfigure}
     \begin{subfigure}[h]{0.32\linewidth}
     \centering
     \includegraphics[width=\textwidth]{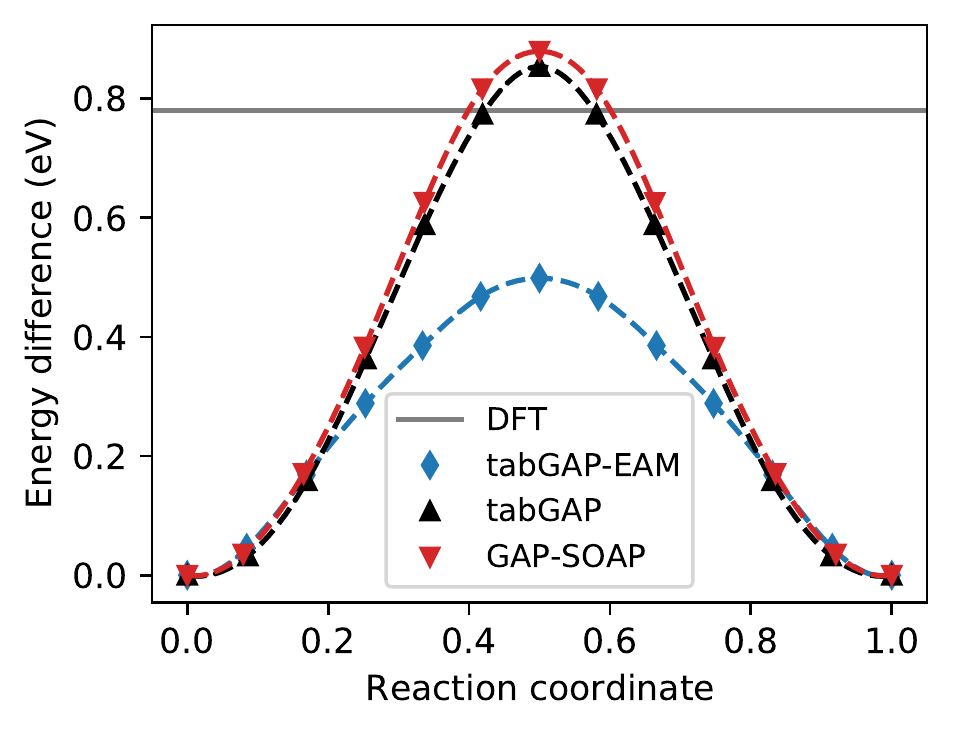}
     \includegraphics[width=0.9\textwidth]{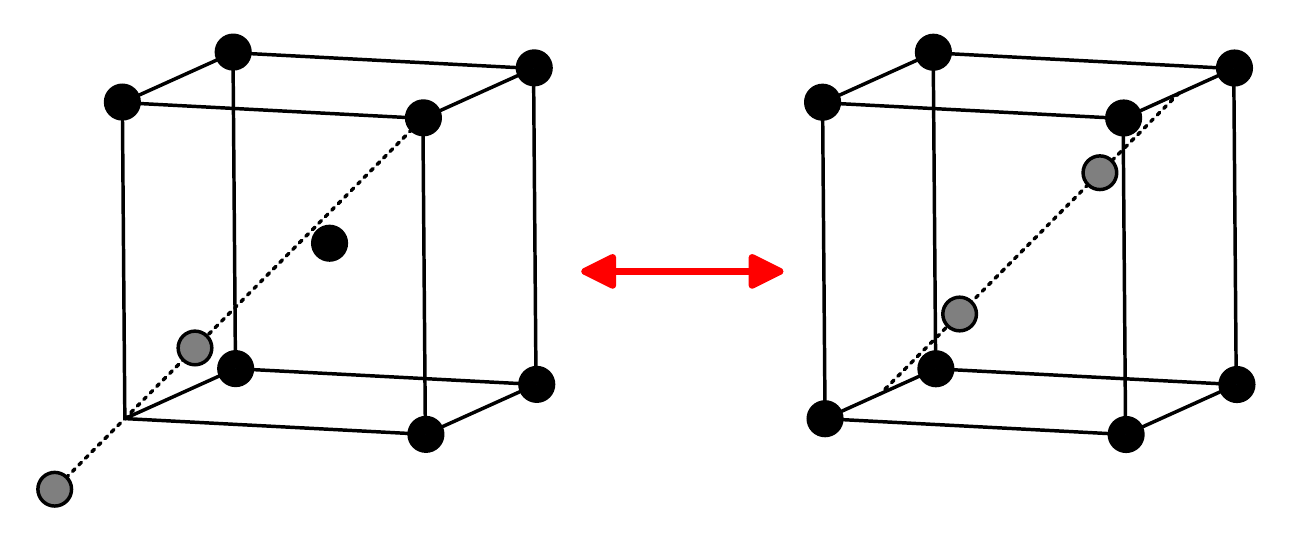}
     \caption{1NN translation \hkl[110] $\rightarrow$ \hkl[110].}
    \end{subfigure}
    \begin{subfigure}[h]{0.32\linewidth}
     \centering
     \includegraphics[width=\textwidth]{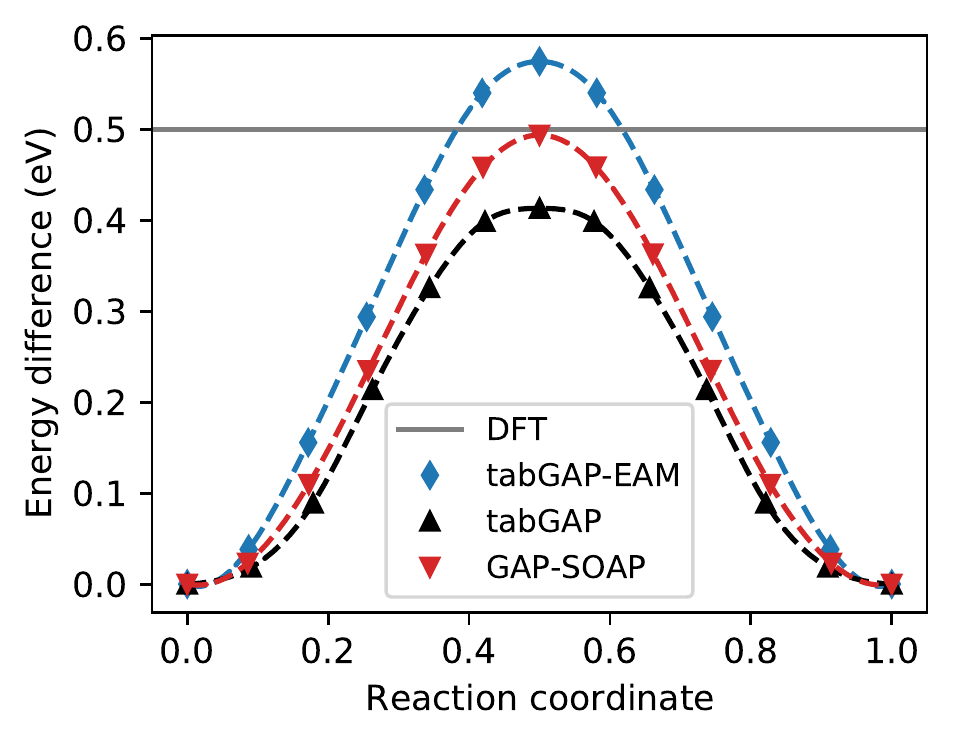}
     \includegraphics[width=0.9\textwidth]{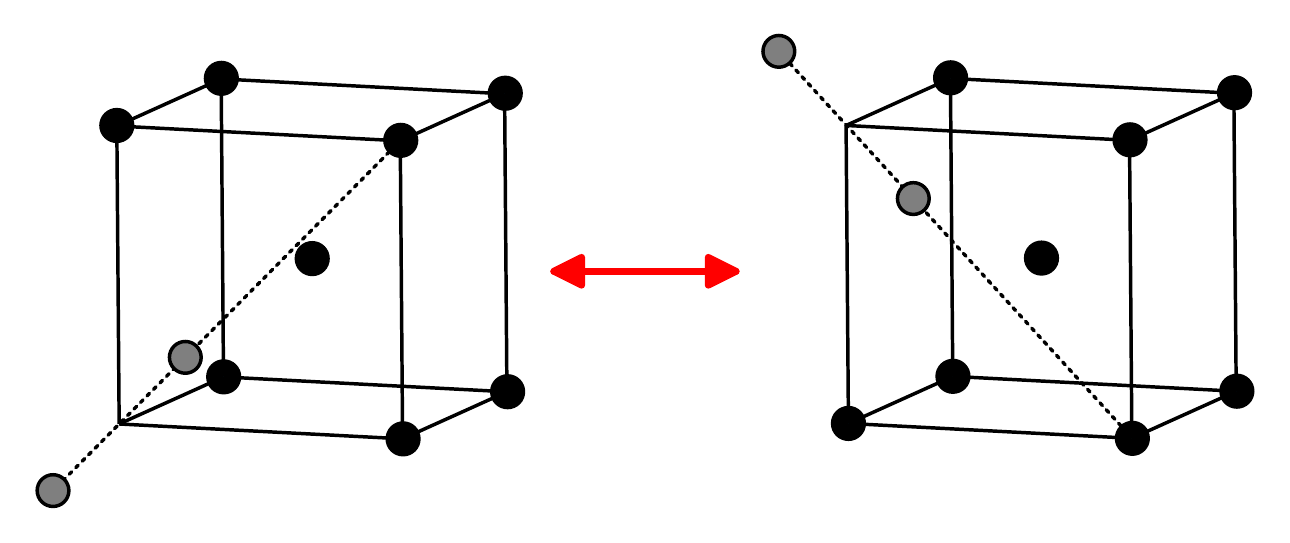}
     \caption{2NN jump \hkl[110] $\rightarrow$ \hkl[-101].}
    \end{subfigure}
    \begin{subfigure}[h]{0.32\linewidth}
     \centering
     \includegraphics[width=\textwidth]{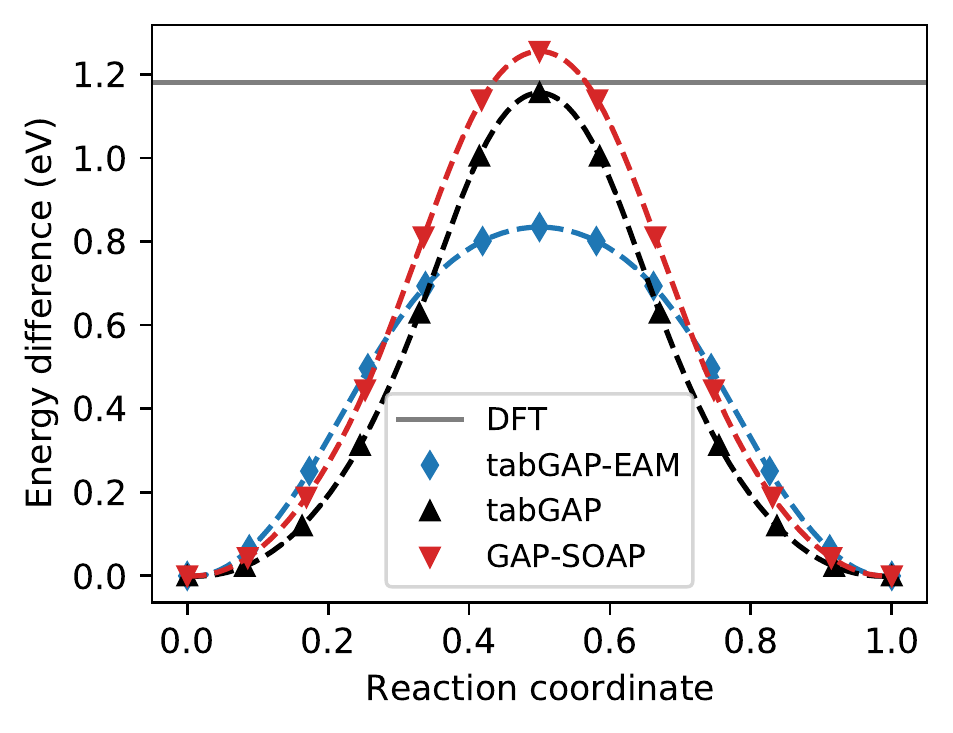}
     \includegraphics[width=0.9\textwidth]{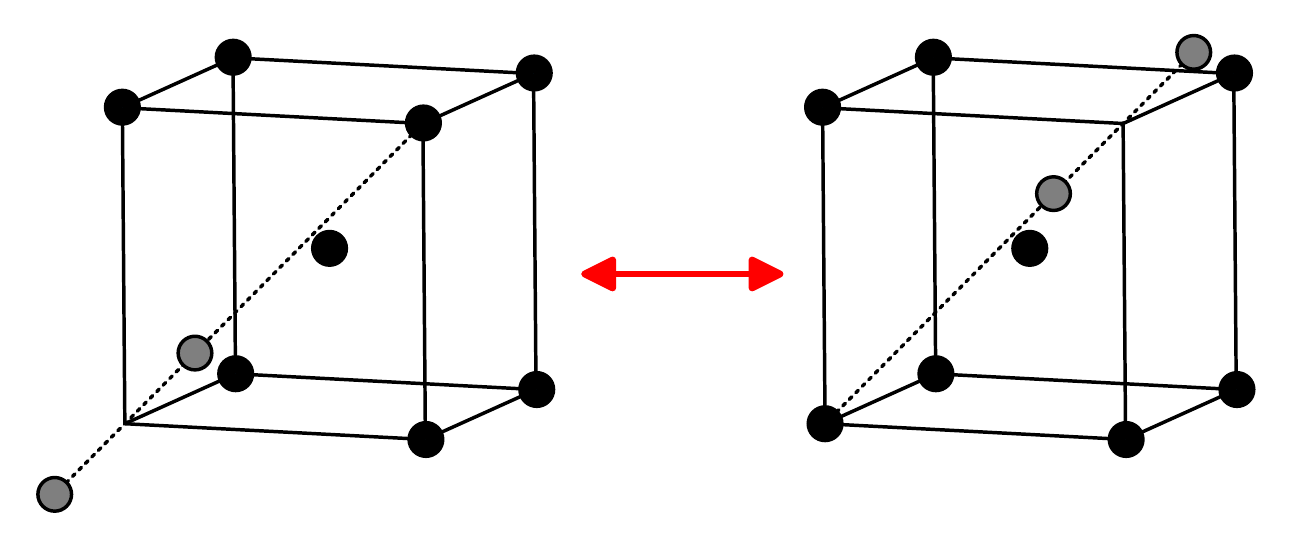}
     \caption{3NN translation \hkl[110] $\rightarrow$ \hkl[110].}
    \end{subfigure}
    \caption{Self-interstitial migration paths and barriers.}
    \label{fig:barriers_sia}
\end{figure}

\begin{figure}[h]
    \centering
    \begin{subfigure}[h]{0.45\linewidth}
     \centering
     \includegraphics[width=\textwidth]{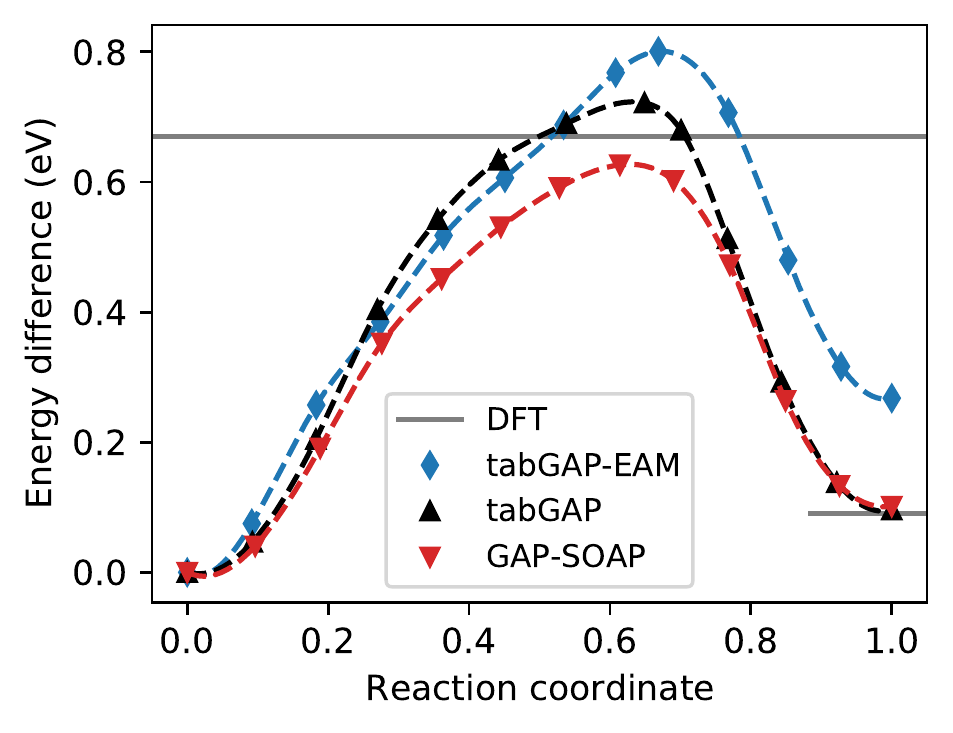}
     \includegraphics[width=0.8\textwidth]{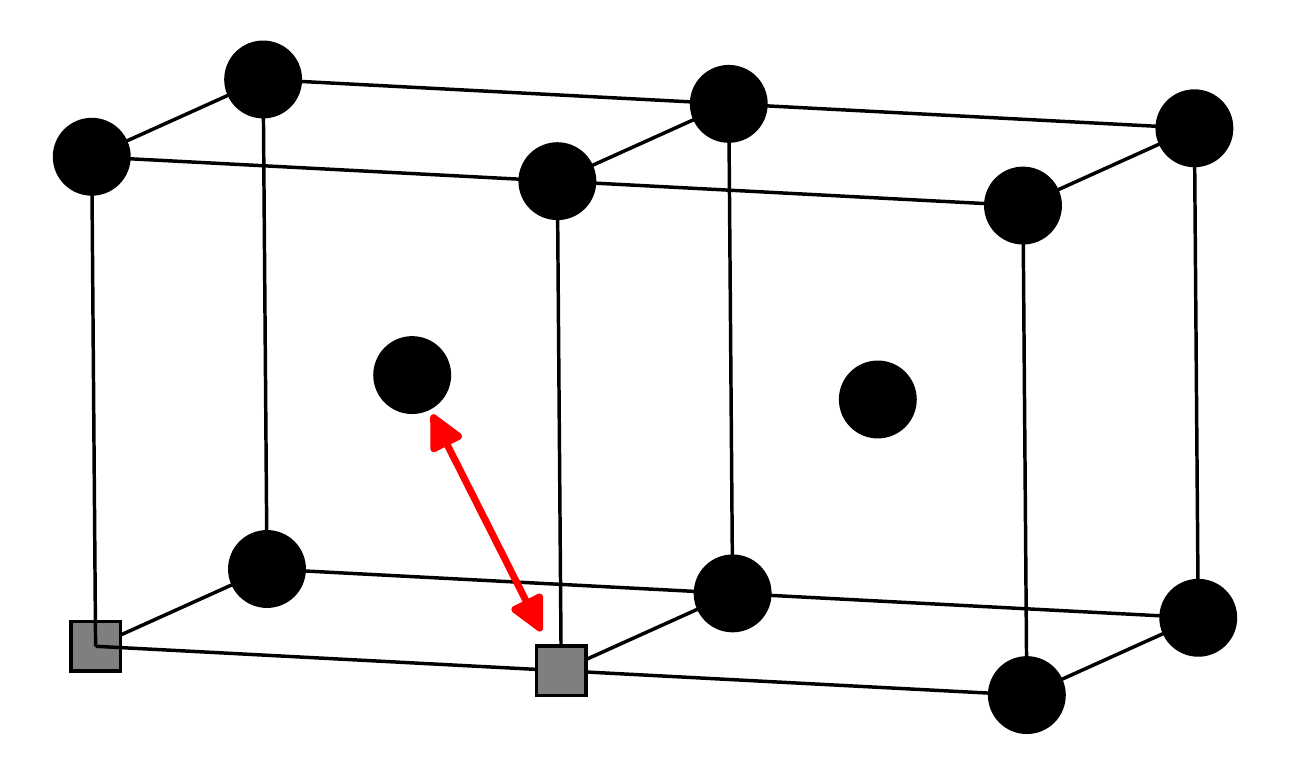}
     \caption{2NN $\xleftrightarrow{}$ 1NN.}
    \end{subfigure}
    \begin{subfigure}[h]{0.45\linewidth}
     \centering
     \includegraphics[width=\textwidth]{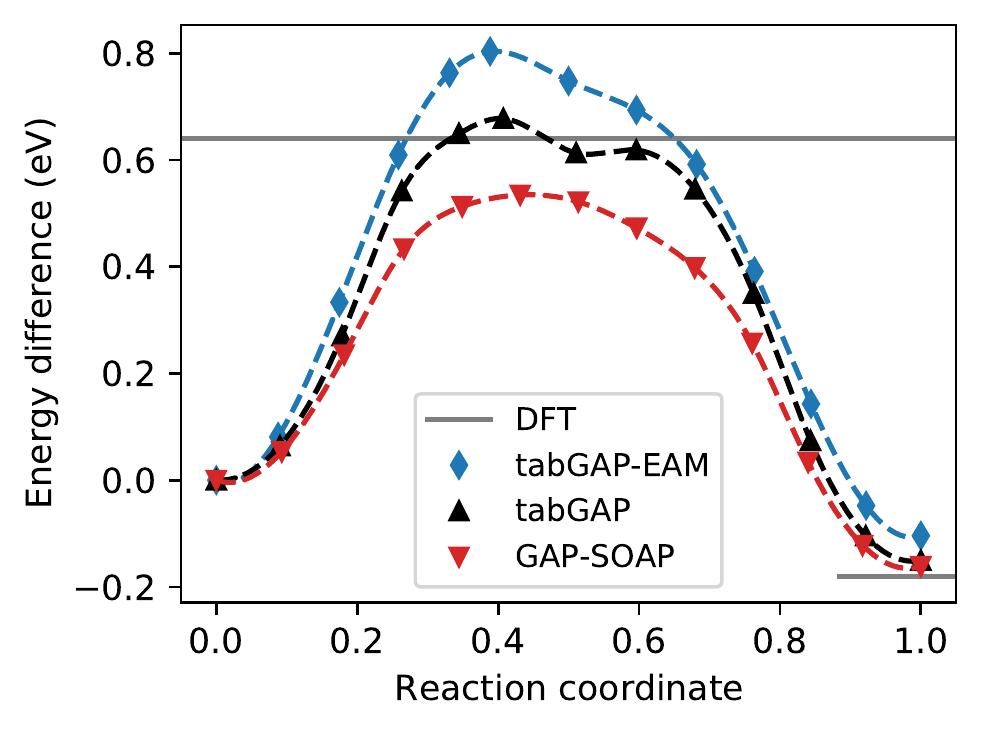}
     \includegraphics[width=0.8\textwidth]{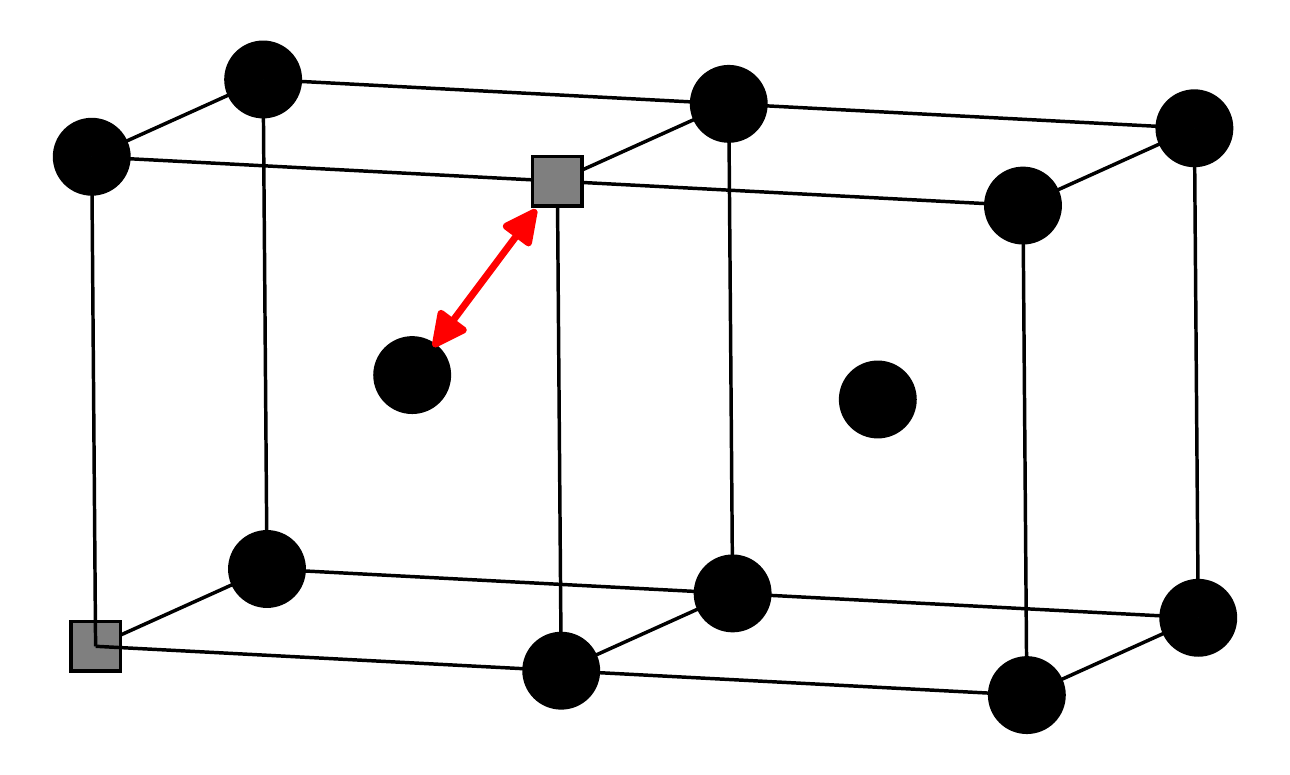}
     \caption{3NN $\xleftrightarrow{}$ 1NN.}
    \end{subfigure}
    \begin{subfigure}[h]{0.45\linewidth}
     \centering
     \includegraphics[width=\textwidth]{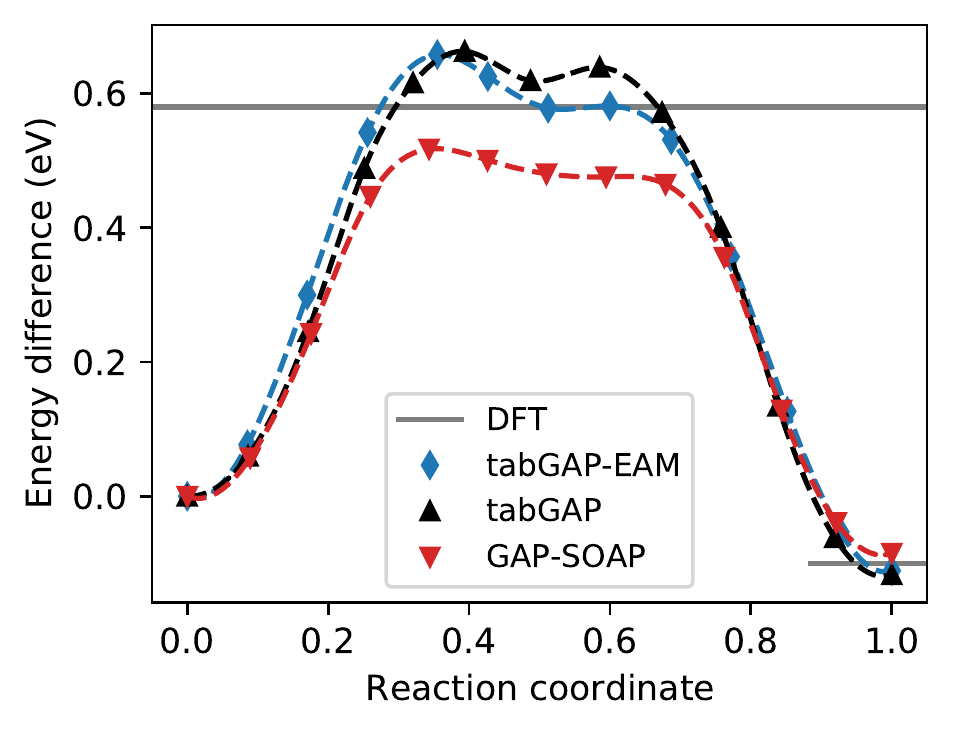}
     \includegraphics[width=0.8\textwidth]{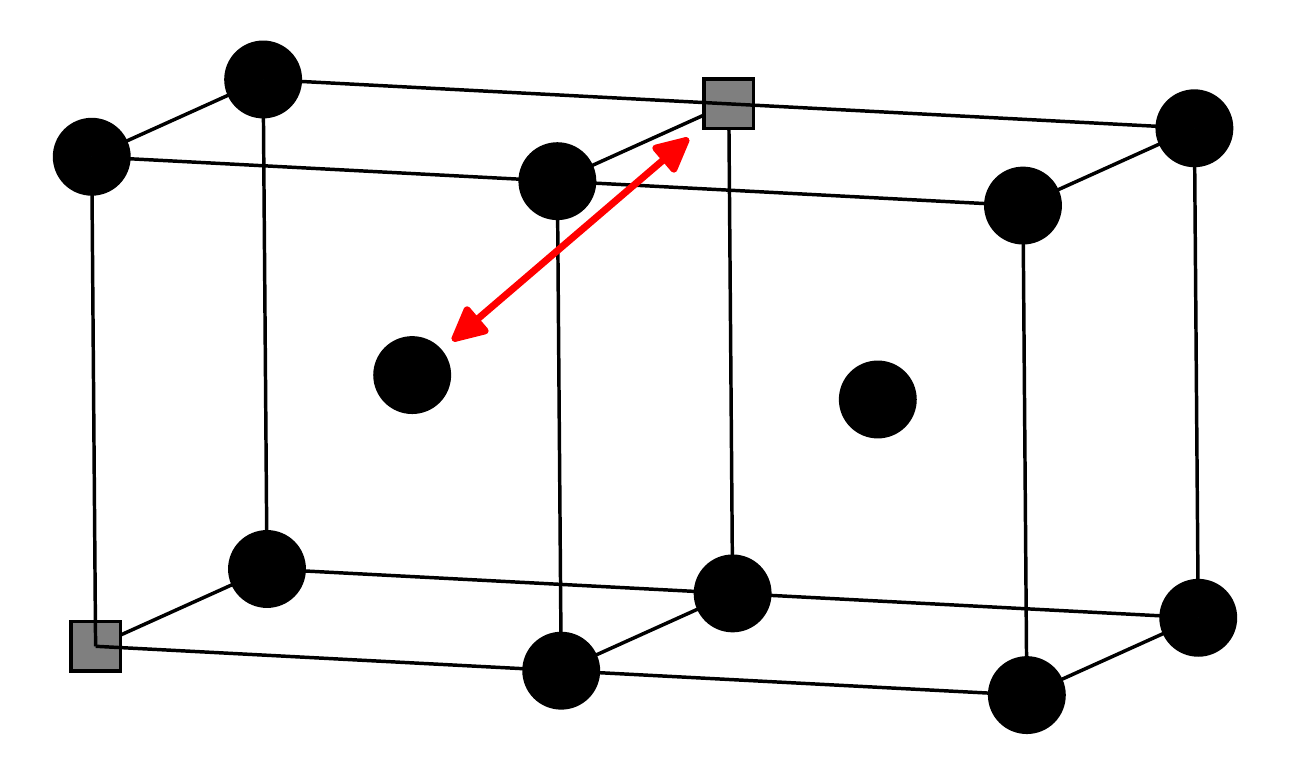}
     \caption{5NN $\xleftrightarrow{}$ 1NN.}
    \end{subfigure}
    \begin{subfigure}[h]{0.45\linewidth}
     \centering
     \includegraphics[width=\textwidth]{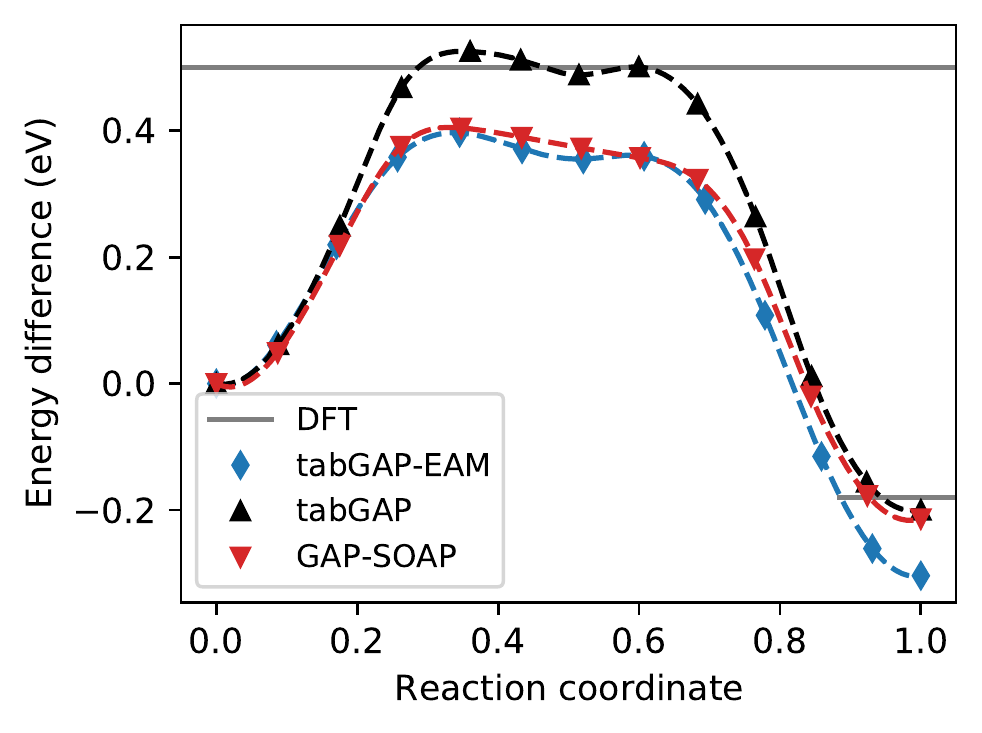}
     \includegraphics[width=0.8\textwidth]{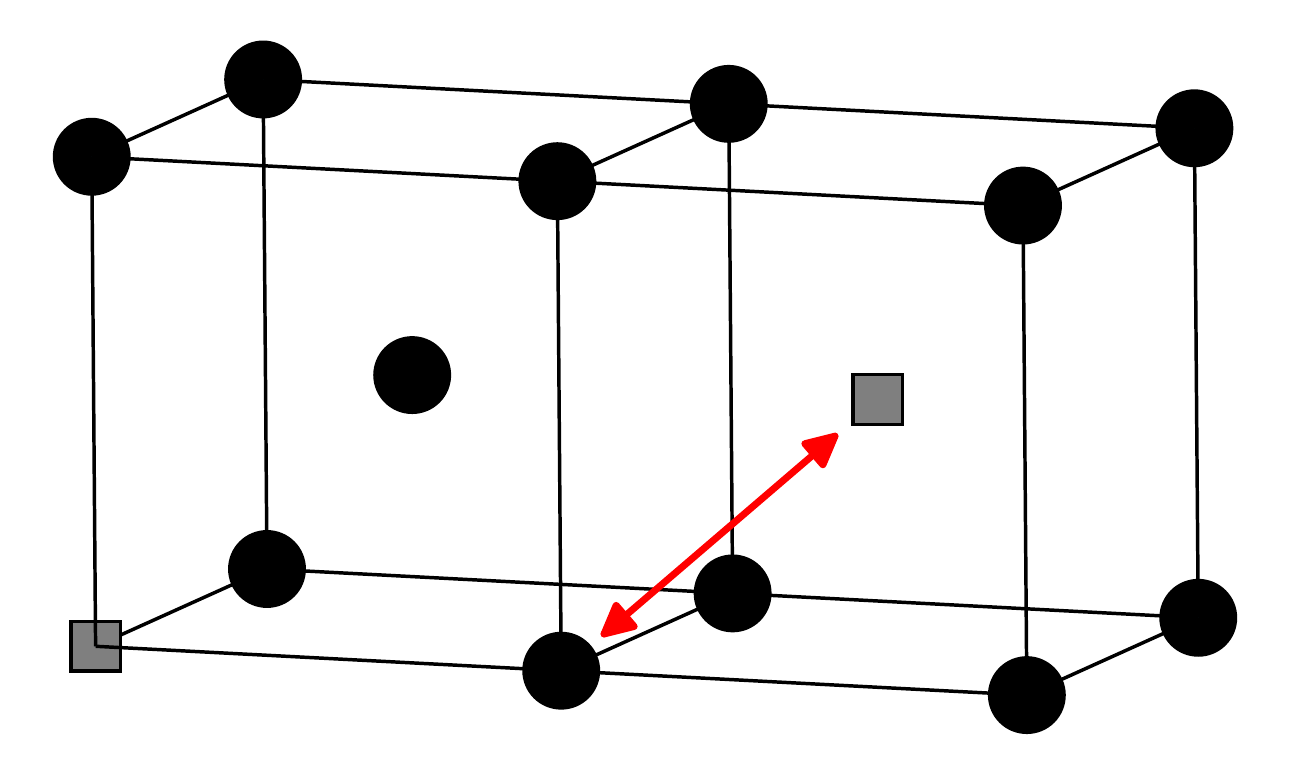}
     \caption{4NN $\xleftrightarrow{}$ 2NN.}
    \end{subfigure}
    \caption{Divacancy migration paths and barriers.}
    \label{fig:barriers_divac}
\end{figure}

Figures~\ref{fig:barriers_sia} and~\ref{fig:barriers_divac} shows the self-interstitial and divacancy migration paths and barriers that are plotted in Fig. 9 in the main text. The migration energies are listed in Tables~\ref{tab:sia} and~\ref{tab:divac}. The DFT values are from Ref.~\cite{malerba_comparison_2010}.

\FloatBarrier
\begin{table}[h]
    \centering
    \caption{Migration energies (eV) for the self-interstitial migration paths shown in Fig.~\ref{fig:barriers_sia}.}
    \begin{tabular}{lrrrr}
     \toprule
      Path & DFT & tabGAP-EAM & tabGAP & GAP-SOAP \\
      \midrule
        Main path & 0.34 & 0.29 & 0.30 & 0.31 \\
        On-site rotation \hkl[101] $\rightarrow$ \hkl[011] & 0.56 & 0.42 & 0.57 & 0.62 \\
        On-site rotation \hkl[110] $\rightarrow$ \hkl[111] & 0.76 & 0.47 & 0.81 & 0.89 \\
        1NN translation \hkl[110] $\rightarrow$ \hkl[110] & 0.78 & 0.50 & 0.85 & 0.88 \\
        2NN jump \hkl[110] $\rightarrow$ \hkl[-101] & 0.50 & 0.57 & 0.41 & 0.49 \\
        3NN translation \hkl[110] $\rightarrow$ \hkl[110] & 1.18 & 0.83 & 1.16 & 1.26 \\
        \bottomrule
    \end{tabular}
    \label{tab:sia}
\end{table}

\begin{table}[h]
    \centering
    \caption{Migration energies (eV) for the forward/backward divacancy migration jumps shown in Fig.~\ref{fig:barriers_divac}.}
    \begin{tabular}{lrrrr}
     \toprule
      Path & DFT & tabGAP-EAM & tabGAP & GAP-SOAP \\
      \midrule
        2NN $\xleftrightarrow{}$ 1NN & 0.67/0.58 & 0.80/0.53 & 0.72/0.63 & 0.63/0.52 \\
        3NN $\xleftrightarrow{}$ 1NN & 0.64/0.82 & 0.80/0.91 & 0.68/0.83 & 0.53/0.70 \\
        5NN $\xleftrightarrow{}$ 1NN & 0.58/0.68 & 0.66/0.77 & 0.66/0.78 & 0.52/0.60 \\
        4NN $\xleftrightarrow{}$ 2NN & 0.50/0.68 & 0.40/0.70 & 0.53/0.73 & 0.40/0.62 \\
        \bottomrule
    \end{tabular}
    \label{tab:divac}
\end{table}

\bibliography{mybib}